\documentclass[%
 aip,
 amsmath,amssymb,
 reprint,%
]{revtex4-1}

\usepackage{graphicx}
\usepackage{dcolumn}
\usepackage{bm}
\usepackage{xspace}
\usepackage{placeins}

\usepackage{xspace}
\usepackage{xcolor}

\renewcommand{\v}[1]{\ensuremath{\mathbf{#1}}}
\newcommand{\rtt}[1]{\textcolor{black}{#1}}
\newcommand{\btt}[1]{\textcolor{black}{#1}}

\begin{document}

\clearpage

\title{Comparing first-principles density functionals plus corrections for the lattice dynamics of YBa$_2$Cu$_3$O$_6$}


\author{Jinliang Ning}
\affiliation{Department of Physics and Engineering Physics, Tulane University, New Orleans, Louisiana 70118, USA}

\author{Christopher Lane}
\affiliation{Theoretical Division, Los Alamos National Laboratory, Los Alamos, New Mexico 87545, USA}

\author{Bernardo Barbiellini}
\affiliation{Department of Physics, School of Engineering Science, LUT University, FI-53851 Lappeenranta, Finland}
\affiliation{Department of Physics, Northeastern University, Boston, Massachusets 02115, USA}

\author{Robert S. Markiewicz}
\affiliation{Department of Physics, Northeastern University, Boston, Massachusets 02115, USA}

\author{Arun Bansil}
\affiliation{Department of Physics, Northeastern University, Boston, Massachusets 02115, USA}

\author{Adrienn Ruzsinszky}
\email{aruzsin@tulane.edu}
\affiliation{Department of Physics and Engineering Physics, Tulane University, New Orleans, Louisiana 70118, USA}

\author{John P. Perdew}
\email{perdew@tulane.edu}
\affiliation{Department of Physics and Engineering Physics, Tulane University, New Orleans, Louisiana 70118, USA}

\author{Jianwei Sun}
\email{jsun@tulane.edu}
\affiliation{Department of Physics and Engineering Physics, Tulane University, New Orleans, Louisiana 70118, USA}

\date{\today}

\begin{abstract}
The enigmatic mechanism underlying unconventional high-temperature superconductivity, especially the role of lattice dynamics, has remained a subject of debate. Theoretical insights have long been hindered due to the lack of an accurate first-principles description of the lattice dynamics of cuprates. Recently, using the r2SCAN meta-GGA functional, we were able to achieve accurate phonon spectra of an insulating cuprate YBa$_2$Cu$_3$O$_6$, and discover significant magnetoelastic coupling in experimentally interesting Cu-O bond stretching optical modes [Ning \textit{et al.}, Phys. Rev. B 107, 045126 (2023)]. We extend this work by comparing PBE and r2SCAN performances with corrections from the on-site Hubbard U and the D4 van der Waals (vdW) methods, aiming at further understanding on both the materials science side and the density functional side. We demonstrate the importance of vdW and self-interaction corrections for accurate first-principles YBa$_2$Cu$_3$O$_6$ lattice dynamics. Since r2SCAN by itself partially accounts for these effects, the good performance of r2SCAN is now more fully explained. In addition, the performances of the Tao-Mo series of meta-GGAs, which are constructed in a different way from SCAN/r2SCAN, are also compared and discussed.


\end{abstract}

\maketitle


\section{\label{sec:level1}Introduction}
Despite the decades of vigorous efforts devoted to the understanding of unconventional high-temperature superconductivity in the cuprates, a consensus on the underlying mechanism has yet to be reached \cite{1986HTc,RVBAnderson,pairRMP,arpsRMP,keimer2015,Sobota2021}. Early theoretical works \cite{bcsnotwork,giustino08nature,YBCO7LDA} suggested that the conventional BCS theory (electron-phonon coupling mechanism) \cite{BCS1,BCS2,BCS3} could not account for such high critical temperatures in cuprate superconductors. However, a more intricate and intriguing picture has been suggested by recent experimental findings\cite{ELmixHTc,rezniknatureEPC,natureEPC,NCCOanomaly, LSCOanomaly,rezniknatureEPC,isotopeHTc,B2212EPC}. Strong anomalies in Cu-O bond-stretching modes are found near optimal doping, which is associated with charge inhomogeneity and beyond previous pictures and understandings \cite{rezniknatureEPC}. Optical spectroscopy results indicate that the electron-phonon coupling contributes at least 10\% \btt{of} the bosonic pairing glue, although antiferromagnetic spin fluctuations are deemed as the main mediators \cite{dal2012}. Moreover, the electronic interactions and the electron-phonon coupling are found to reinforce each other in a positive-feedback loop, which in turn enhances superconductivity, as suggested by recent ARPES observations \cite{B2212EPC}. 

Part of the reason why the role of phonons was dismissed by the theoretical community was that previous density functional theory (DFT) calculations at the local spin density approximation (LSDA) and generalized gradient approximation (GGA) levels failed to find strong electron-phonon coupling in related cuprates\cite{giustino08nature}. This issue is related to and compounded by the fact that these density functional approximations (DFAs) cannot stabilize the correct electronic and magnetic ground state in the parent phase, let alone its evolution with doping\cite{yubostripe,James_SCAN_Cuprates,Chris_LCO}. While corrections such as the Hubbard U \cite{HubbardU,linearU,LDAU_MnO,LSDAU_NiO} method can stabilize the antiferromagnetic (AFM) ground state \cite{sterling2021}, their structural predictions can be unexpected and uncontrollable \cite{jarlborg2014}. Obviously, an {\it ab initio} treatment is required to capture simultaneously the electronic, magnetic and lattice degrees of freedom.

Recently, utilizing the r2SCAN meta-GGA functional \cite{r2SCAN}, some of us\cite{YBCO6_NJL} have been able to stabilize the AFM state of the pristine oxide YBa$_2$Cu$_3$O$_6$, and faithfully reproduce the experimental phonon dispersions. We further found significant magnetoelastic coupling in numerous high-energy Cu-O bond stretching optical branches, where the AFM results improve over the soft nonmagnetic phonon bands\cite{YBCO6_NJL}. Moreover, these phonons correspond to breathing modes within the CuO$_2$ plane, suggesting a sensitive dependence on magnetoelastic coupling, which may facilitate a positive-feedback loop between electronic, magnetic, and lattice degrees of freedom. The r2SCAN functional is a modified and improved version of the strongly-constrained and appropriately-normed (SCAN) meta-GGA functional \cite{SCAN,SCAN_NChem}, which satisfies 17 exact constraints, \rtt{and} has demonstrated an excellent performance across a diverse range of bonding environments. For cuprates, SCAN accurately predicts the correct half-filled AFM ground state and the observed insulator-metal transition upon doping \cite{yubostripe,James_SCAN_Cuprates}. Moreover, SCAN provides improved estimates of lattice constants, across correlated and transition metal compounds \cite{SCAN,SCAN_NChem, yubostripe, James_SCAN_Cuprates, yubo_MO, Yubo_TiO2, Peng_MnO2, Peng_MO, ning_MBT,SIO_Chris, Kanun_NPJ,SmB6}. Thus, SCAN is promising in accurate descriptions of lattice dynamics of cuprates and associated electron-phonon couplings, by virtue of its ability to capture the electronic and magnetic ground states. \rtt{Unfortunately, reliable phonon spectra from SCAN calculations can be challenging due to numerical instability problems, although \rtt{they could still be feasible} with extra computational cost \cite{r2SCAN_phonon,r2SCAN_rVV10} or more advanced phonon calculation techniques \cite{GQHA1,GQHA2,phonon_irr_der}.} By design, r2SCAN \cite{r2SCAN} solves the numerical instability problem and delivers accurate, transferable, and reliable lattice dynamics for various systems with different bonding characteristics\cite{r2SCAN_phonon}. We thus chose r2SCAN instead of SCAN for the study of lattice dynamics of YBa$_2$Cu$_3$O$_6$ and achieved remarkable success.  

Despite this success, there still exists a notable residual softening trend in the Cu-O bond-stretching optical phonon branches from r2SCAN, especially in the full-breathing modes, for which we achieved further improvements when a Hubbard U correction is applied to r2SCAN \cite{YBCO6_NJL}. \btt{Note that similar improvements from DFT+U \btt{or averaged correction for self-interaction error\cite{PZSIC}} were reported for optical modes of \btt{Mott insulators} La$_2$CuO$_4$\cite{sterling2021,model_Falter1} and UO$_2$\cite{UO2_phonon}.} \btt{The good performance of r2SCAN/SCAN on cuprates has been ascribed to the power of satisfying exact constraints by design in SCAN/r2SCAN\cite{SCAN,r2SCAN,r4SCAN,Aaron_review} and the reduction of self-interaction error\cite{James_SCAN_Cuprates,Kanun_NPJ}. However, previous} studies suggest that vdW corrections are important for first-principles prediction of lattice constants and cohesive energy of ionic solids and heavy metals \cite{vdw_solids_Tao}. In addition, combining vdW correction and self-interaction correction (SIC) is of critical importance for ground state electronic, structural and energetic properties of transition metal monoxides \cite{Peng_MO}. Therefore, it is expected that the vdW correction and its combination with SIC are crucial for obtaining accurate phonon dispersions of cuprates based on DFT. 


To confirm this, in this work we extend our lattice dynamics study of YBa$_2$Cu$_3$O$_6$ by comparing the PBE and r2SCAN performances with corrections from the Hubbard U (applied to the $d$ orbitals of Cu) and the D4 van der Waals (vdW) correction methods \cite{D4,D4_test,D4_parameter,r2SCAN_D4}, aiming at further understanding both the physics of cuprate lattice dynamics and density functionals. We demonstrate the importance of vdW interactions and SIC for YBa$_2$Cu$_3$O$_6$ lattice dynamics. Since r2SCAN by itself provides a partial account of these effects to a greater degree than PBE, the better performance of r2SCAN is more fully explained. In addition, the performances of the Tao-Mo family of meta-GGAs (TMs) \cite{TM,revTM,rregTM} are also compared and discussed. The original Tao-Mo meta-GGA (TM) \cite{TM} is constructed based on a density matrix expansion of the exchange hole model, while revTM \cite{revTM} and rregTM \cite{rregTM} are two successors with modifications. The revTM includes a correlation correction obtained from the full high-density second-order gradient expansion, while rregTM includes a regulation to the order-of-limit problem\cite{OOL_TM}, paired with a one-electron self-interaction-free correlation energy functional. In comparison, SCAN and r2SCAN are constructed by satisfying exact-constraints on the exchange-correlation energy\cite{SCAN,r2SCAN,r4SCAN,Aaron_review}. Due to the inherently different way the Tao-Mo meta-GGAs are constructed compared to SCAN/r2SCAN, a comparison of their performances will be interesting and is expected to shed light on both the materials science side in cuprate lattice dynamics and the DFT side. Due to the absence of D4 parametrizations matched to TM functionals, the effects of vdW corrections to TMs are not considered in this work.

The synergy of long-range vdW corrections and +U SIC that we find here for cuprate lattice dynamics has also been found\cite{Peng_MO} for structural properties and structural phase transitions in MnO, FeO, CoO, and NiO.

\section{\label{sec:level1}Methods}
First-\btt{principles} calculations were performed using the pseudopotential projector-augmented wave method\cite{PAW,PAW_vasp} with the Vienna {\it ab initio} simulation package (VASP) \cite{VASP,VASP2} with an energy cutoff of 600 eV for the plane-wave basis set. Several exchange-correlation functionals including PBE at the GGA level, and r2SCAN \cite{r2SCAN, r2SCAN_phonon}, TM\cite{TM}, revTM\cite{revTM}, and rregTM\cite{rregTM} at the meta-GGA level were used. For the D4 vdW correction, we use the literature parametrizations fitted separately for PBE ($s\mathrm{_6}$=1.0000, $s\mathrm{_8}$=0.9595, $a\mathrm{_1}$=0.3857, $a\mathrm{_2}$=4.8069) \cite{D4_parameter} and for r2SCAN ($s\mathrm{_6}$=1.0000, $s\mathrm{_8}$=0.6019, $a\mathrm{_1}$=0.5156, $a\mathrm{_2}$=5.7734)\cite{r2SCAN_D4}. A Gamma-centered 8$\times$8$\times$4 mesh for the $\textbf{k}$-space sampling is used for the relaxation of the unit cell with a G-type AFM structure, while a 2$\times$2$\times$2 mesh is used for the 2$\times$2$\times$1 supercells for force constant calculations of the phonon dispersion. \rtt{The Fermi smearing method (ISMEAR = -1) was used, set with a low electronic temperature (SIGMA = 0.002 eV).} All atomic sites in the unit cell along with the cell dimensions were relaxed using a conjugate gradient algorithm to minimize the energy with an atomic force tolerance of 0.001 eV/\AA~ and a total energy tolerance of $10^{-7}$ eV. The harmonic force constants were extracted from VASP using the finite displacement method (0.015\AA) as implemented in the Phonopy code \cite{phonopy}.

\section{\label{sec:level1}Results}

\begin{figure}[ht!]
\centering
\includegraphics[width=\linewidth]{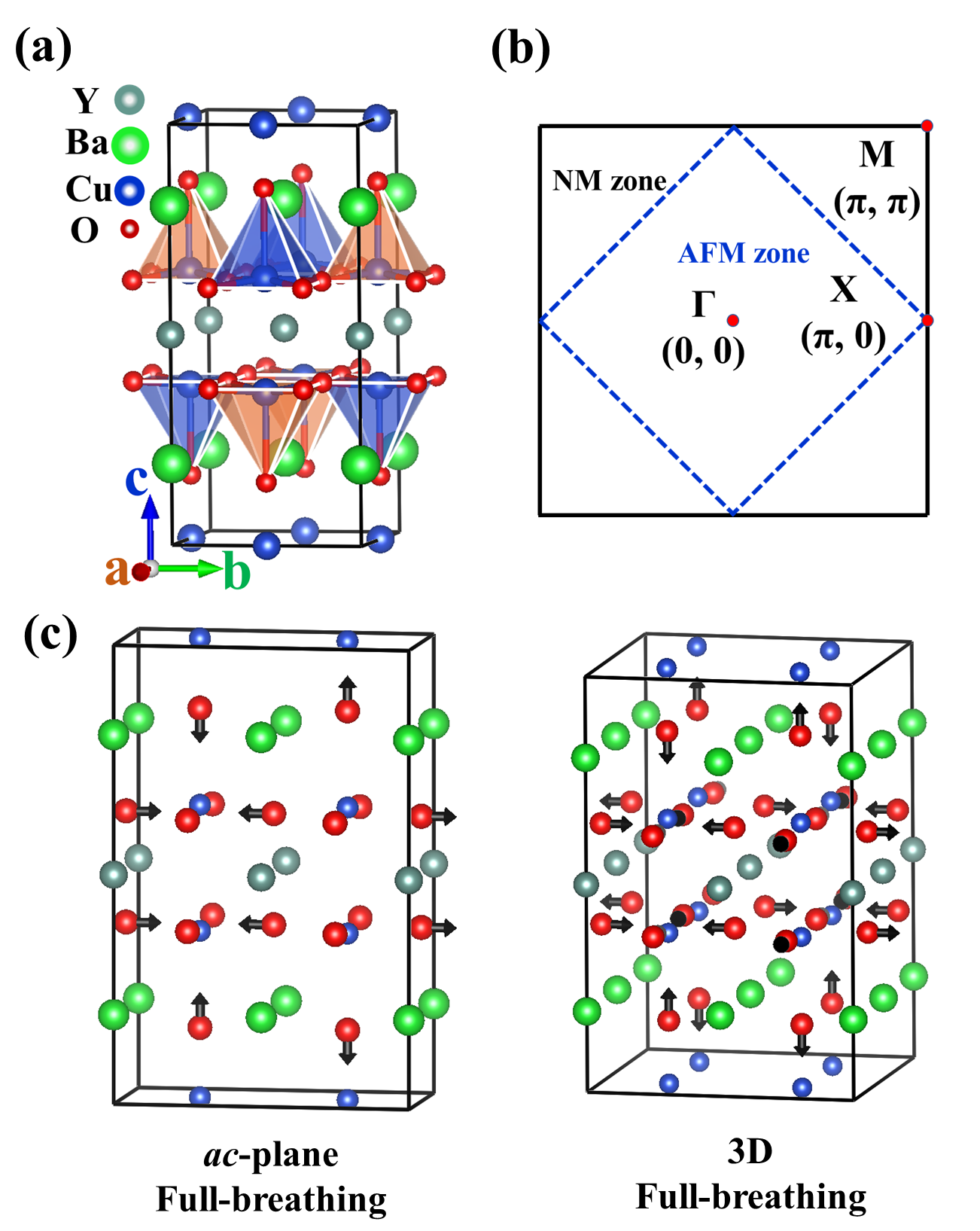}
\caption{(a) Crystal structure of YBa$_2$Cu$_3$O$_6$ where the related G-type AFM structure is highlighted by coloring the corner-sharing Cu-O pyramids blue(orange) for spin up(spin down). \btt{Note that the chain Cu atoms not forming Cu-O pyramids are nonmagnetic.} (b) A schematic of the nonmagnetic (black solid line) and G-type AFM (blue dashed line) Brillouin zones and high-symmetry $\textbf{k}$-points. (c) Schematic of the typical $ac$-plane and 3D full-breathing modes.}\label{fig:demo}
\end{figure} 

\begin{table*}[htbp]
\caption{Calculated lattice constants, volume, Cu magnetic moment, Cu-O plane buckling angle $\angle$O-Cu-O, and Cu-O bond lengths for YBa$_2$Cu$_3$O$_6$ in the G-type AFM phase, along with the available experimental data. The Cu-O bond length for the two adjacent Cu-O planes in the $ab$ plane, where Cu shows local magnetic moment $m$, is denoted by $d_{\mathrm{Cu-O}}$, while $z_{\mathrm{Cu-O_{ap}}}$ (nonmagnetic Cu-apical O bond) and $z^{\prime}_{\mathrm{Cu-O_{ap}}}$ (magnetic Cu-apical O bond) denote the Cu-O bond lengths along the $c$ direction. For PBE and r2SCAN, the Hubbard U and vdW (D4 and rVV10) corrections are considered. The choice of Hubbard U values is guided by both experimental lattice constants and Cu magnetic moment.}
\label{tab:latticeU}
\begin{ruledtabular}
\begin{tabular}{lllllllllll}
DFA & U & vdW &	$a$	(\AA)&	$c$ (\AA)	&	$V$	({\AA}$^3$) &	$m$ ($\mu_B$)	&	$d_{\mathrm{Cu-O}}$ (\AA)	&	$\angle$O-Cu-O ($^\circ$)	&	$z_{\mathrm{Cu-O_{ap}}} $ (\AA)	&	$z^{\prime}_{\mathrm{Cu-O_{ap}}}$ (\AA)	\\
\cline{1-11}
PBE	&	0	&	--	&	3.8819	&	12.1905	&	183.70	&	0.00	&	1.948	&	170.32	&	1.816	&	2.672	\\
	&	6	&	--	&	3.8750	&	12.0379	&	180.76	&	0.59	&	1.950	&	167.06	&	1.811	&	2.551	\\
	&	6	&	D4	&	3.8525	&	11.8702	&	176.17	&	0.59	&	1.938	&	167.25	&	1.803	&	2.492	\\
 \cline{1-11}
r2SCAN	&	0	&	--	&	3.8570	&	11.9417	&	177.65	&	0.45	&	1.937	&	169.28	&	1.805	&	2.554	\\
	&	5	&	--	&	3.8562	&	11.8321	&	175.95	&	0.66	&	1.941	&	167.00	&	1.795	&	2.472	\\
	&	4	&	D4	&	3.8485	&	11.8032	&	174.82	&	0.62	&	1.936	&	167.41	&	1.794	&	2.469	\\
	&	4	&	rVV10	&	3.8482	&	11.7638	&	174.21	&	0.62	&	1.936	&	167.41	&	1.793	&	2.451	\\
 \cline{1-11}
TM	    &	0	&	--	&	3.8651	&	11.8224	&	176.61	&	0.15 (0.3)	&	1.939	&	170.31	&	1.812	&	2.502	\\
	&	5	&	--	&	3.8626	&	11.7186	&	174.84	&	0.62	&	1.943	&	167.56	&	1.803	&	2.421	\\
revTM	&	0	&	--	&	3.8649	&	11.8952	&	177.68	&	0.09 (0.3)	&	1.940	&	170.09	&	1.810	&	2.535	\\
	&	5	&	--	&	3.8621	&	11.7919	&	175.89	&	0.61	&	1.943	&	167.52	&	1.802	&	2.451	\\
rregTM	&	0	&	--	&	3.8982	&	11.9688	&	181.88	&	0.39	&	1.956	&	170.17	&	1.832	&	2.548	\\
	&	5	&	--	&	3.8941	&	11.8648	&	179.92	&	0.63	&	1.960	&	166.74	&	1.824	&	2.445	\\
	\cline{1-11}
Expt.	&	    &	    &	3.8544$^a$	&	11.8175$^a$	&	175.57$^a$	&	0.55$^b$	&	1.940	&	166.78	&	1.786	&	2.471	\\

\end{tabular}
\end{ruledtabular}
\raggedright 
$^a$Powder neutron diffraction at temperature of 5 K\cite{ybco6_LC_expt}.\\
$^b$Single crystal neutron scattering\cite{casalta1994absence}.
\end{table*}

\begin{figure}[htb]
\centering
\includegraphics[width=\linewidth]{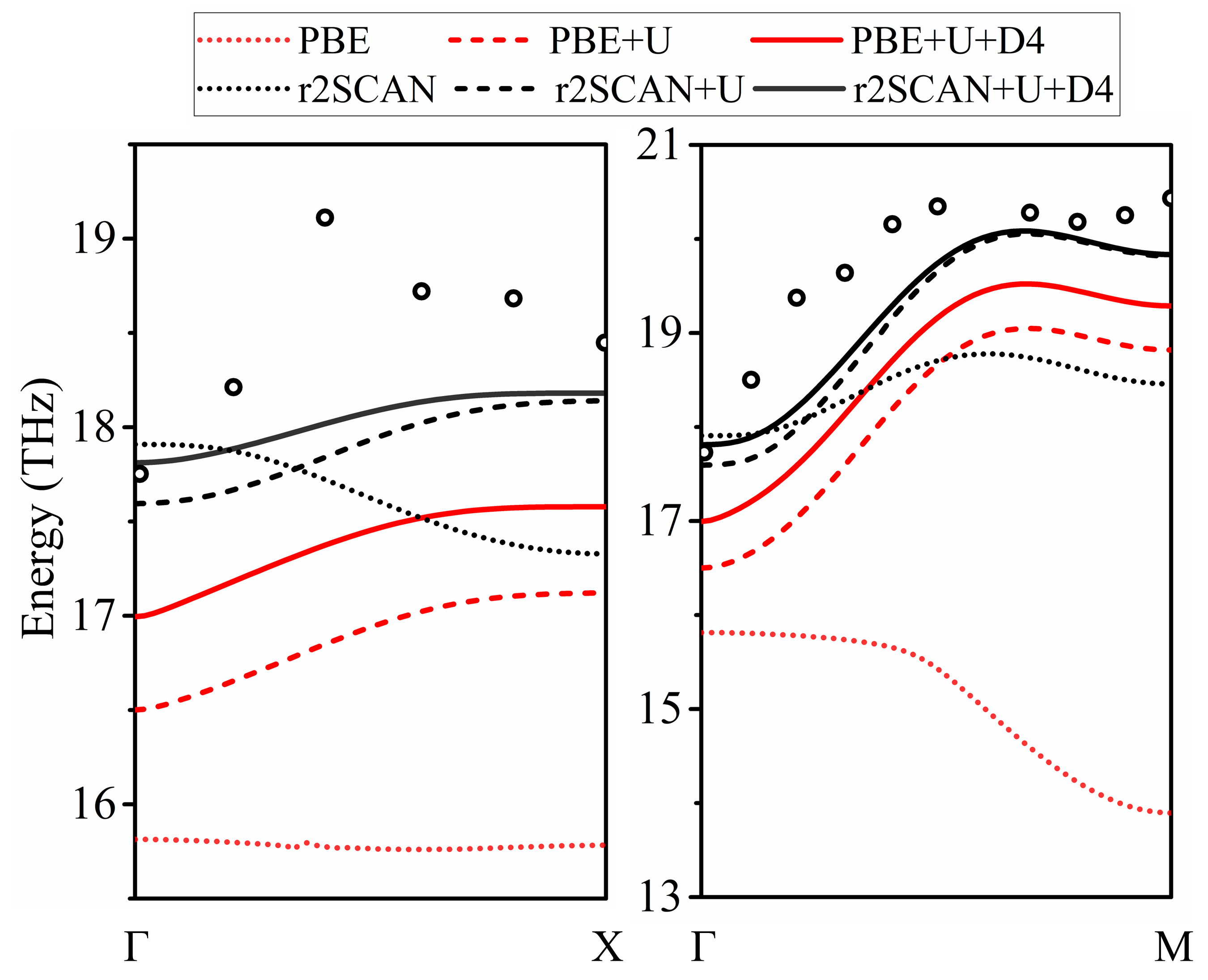}
\caption{Comparison of the $ac$-plane full-breathing branch (highest branch of $\Delta1$, left panel) and the 3D full-breathing branch (highest branch of $\Sigma1$, right panel) for YBa$_2$Cu$_3$O$_6$, calculated from PBE, PBE+U (6 eV), PBE+D4+U (6 eV), r2SCAN, r2SCAN+U (5 eV) and r2SCAN+D4+U (4 eV) methods, with the experimental data (open circles) \cite{YBCO6phonon_expt}. The Brillouin zone and high-symmetry $\textbf{k}$-points are shown in Fig.~\ref{fig:demo}. The phonon results are obtained with supercell interatomic forces calculated with the same DFA (+U+D4) methods as for geometry relaxations, as detailed in Table~\ref{tab:latticeU}.}\label{fig:vdWU}
\end{figure}

\begin{figure}[htbp!]
\centering
\includegraphics[width=\linewidth]{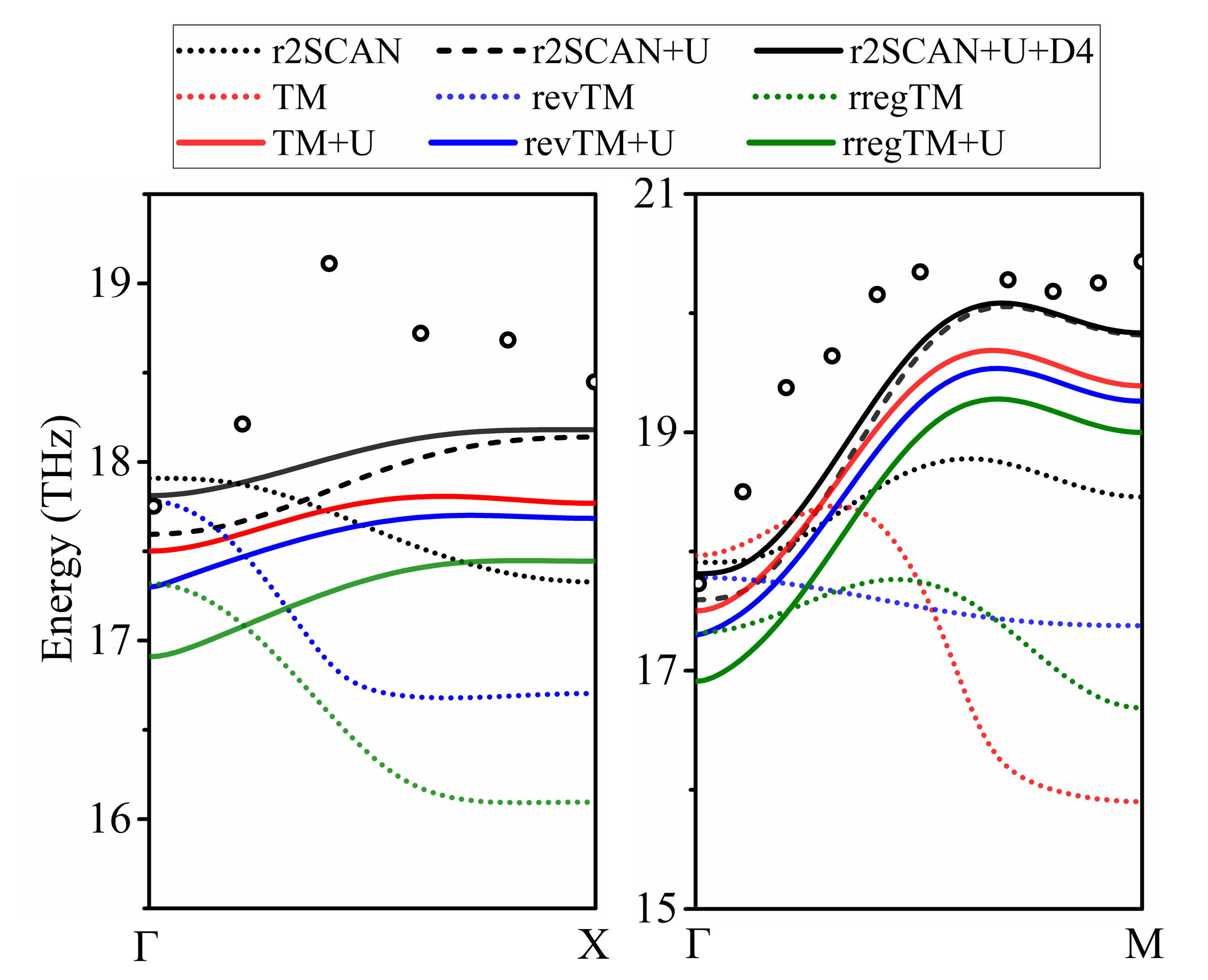}
\caption{Same as Fig.~\ref{fig:vdWU}, but the calculated results are from TM, TM+U (5 eV), revTM, revTM+U (5 eV), rregTM, rregTM+U (5 eV), \rtt{r2SCAN, r2SCAN+U (5 eV), and r2SCAN+D4+U (4 eV)} methods. Due to the complicated band crossing, the bare TM results are not shown for the $ac$-plane full-breathing branch.}\label{fig:TMs}
\end{figure}

\begin{figure*}[htbp!]
\centering
\includegraphics[width=0.8\linewidth]{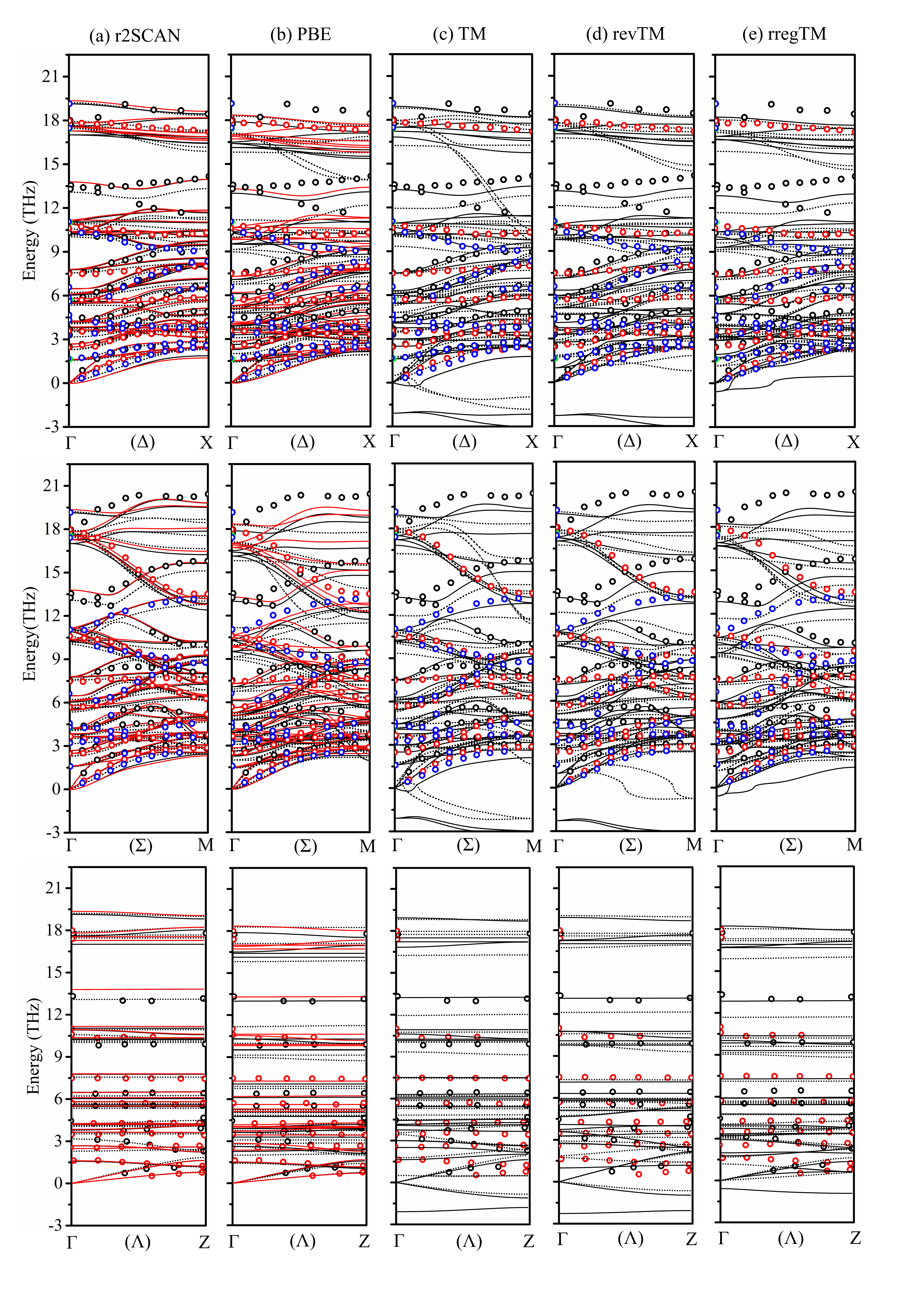}
\caption{\rtt{Comparison of full plots \btt{of} the phonon dispersions of YBa$_2$Cu$_3$O$_6$ including experimental data\cite{YBCO6phonon_expt} (circles) and calculated results from DFAs and corrections based on (a) r2SCAN, (b) PBE, (c) TM, (d) revTM, and (e) rregTM. The bare DFA, DFA+U, and DFA+U+D4 results are represented as black dotted, black solid, and red solid lines, respectively. The U values applied are the same as in Figs.~\ref{fig:vdWU} and ~\ref{fig:TMs}.}} \label{fig:full}
\end{figure*}

Table~\ref{tab:latticeU} presents the comparison of various properties calculated from various methods with available experimental values for YBa$_2$Cu$_3$O$_6$. The bare DFA results are compared with those with Hubbard U and/or vdW corrections. For the bare DFAs considered here, they all overestimate the lattice constants and underestimate the magnetic moments, in different degrees. In particular, PBE overestimates the lattice constants the most and at the same time is not able to stabilize the correct G-type AFM ground state, while r2SCAN and the TMs at the meta-GGA level can stabilize the G-type AFM ground state, and improve over PBE for structural properties. In addition, a notable difference can be observed in the performances of these different meta-GGAs. TM gives the closest structural properties but notably underestimates the magnetic moments. The revTM functional performs similar to TM in structural properties but worse in magnetic moments. The rregTM functional improves in magnetic moments but worsens in structural properties. The r2SCAN functional in general predicts a good combination of structural and magnetic properties. What is more, TM and revTM predict different magnetic moments for the G-type AFM unit cell and the 2$\times$2$\times$1 supercells used for interatomic force constants calculations. This suggests that they could stabilize some spin-density-wave states rather than the simple G-type AFM ground state, \rtt{which can be confirmed to some extent by the imaginary frequencies in the calculated phonon dispersions as shown in  Fig.~\ref{fig:full}. The large spurious imaginary bands from TMs and/or their U-corrected versions also indicate that they may suffer much more serious numerical instability problems compared even to SCAN.} 

Applying a Hubbard U correction to these DFAs will reduce the delocalization error and increase the predicted magnetic moments. At the same time it also improves the structural properties, which will not be true for LSDA+U as we found previously since LSDA already underestimates the lattice constants\cite{YBCO6_NJL}. In general, due to self-interaction reduction, the meta-GGAs require smaller U corrections than PBE\cite{PBESCANU,Yubo_TiO2,Peng_MO}. The structural properties can be further improved with additional vdW corrections, as demonstrated by the PBE+U+D4 and r2SCAN+U+D4 results. Similarly to SCAN, r2SCAN captures intermediate range vdW interactions\cite{SCAN_NChem,r2SCAN_rVV10,Manish_ACS} while PBE captures little\cite{PBE_rVV10}. Therefore, more vdW correction is needed for PBE than for r2SCAN, and correspondingly the structural improvements from the vdW correction are greater when applied to PBE than to r2SCAN. As a comparison and cross-check, almost the same structural and magnetic results are achieved with r2SCAN+rVV10 \cite{r2SCAN_rVV10} and r2SCAN+D4 \cite{r2SCAN_D4}. \rtt{Note that the difference in the predicted lattice constant c from the two vdW corrections is more noticeable. Unfortunately, we are not sure which one is more reliable at the current stage. For the current calculated lattice constants, the zero-point energy effect is not considered. So, only some qualitative trend can be claimed, rather than quantitative comparisons especially when the difference is very small. It is possible that rVV10 is more accurate than D4, but the U value in r2SCAN+U+D4 or r2SCAN+U+rVV10 is too large, \btt{since the resulting magnetic moment is larger than the experimental one.} It is also possible that the b parameter in r2SCAN+rVV10 fitted from the Ar dimer binding curve is overbinding solid systems where the screening effect is usually stronger than that in molecules, as confirmed to some extent by the case of PBE+rVV10L\cite{PBE_rVV10} where a larger b (less vdW correction) is found to work better for layered materials. The current finding of the importance of vdW corrections to property predictions of YBa$_2$Cu$_3$O$_6$} is consistent with previous reports that vdW corrections are important for structural and energetic properties of ionic solids \cite{vdw_solids_Tao}, and that the combination of vdW and Hubbard U corrections is important for the ground state electronic, structural and energetic properties of transition metal monoxides \cite{Peng_MO}. It is also consistent with the underestimation of lattice constants in LSDA, since LSDA tends to overbind weak bonds. Generally speaking, the improvements of meta-GGAs over GGAs can be attributed to the power of satisfying more exact constraints\cite{r4SCAN,Aaron_review}, but, in addition, the self-interaction reduction and better capture of vdW interactions at least contribute to a major part of their improvements. Moreover, for PBE and r2SCAN, although applying larger Hubbard U alone can yield structural properties closer to experiment, it could over-localize {\it d} electrons and predict too large magnetic moments, highlighting the different physics of the two corrections and the necessity of appropriately applying both together. Since the phonon dispersion is determined by the inter-atomic forces, which depend sensitively on the ground state electronic structure and equilibrium atomic positions\cite{r2SCAN_phonon}, the improvements for structural and electronic/magnetic properties from vdW and Hubbard U corrections bode well for more accurate predictions of the lattice dynamics.

The phonon dispersion results for the most challenging and also experimentally most interesting $ac$-plane and three-dimensional (3D) full-breathing branches, as shown in Fig.~\ref{fig:vdWU} and Fig.~\ref{fig:TMs}, confirm our expectations. Both the $ac$-plane and 3D full-breathing modes involve the Cu-O bond-stretching vibrations within the Cu-O plane, and simultaneously the vibration of the apical oxygen in the {\it c} direction. The difference is \btt{that} the $ab$ Cu-O plane bond-stretching vibrations happen along both $a$ and $b$ directions for the 3D full-breathing modes, while only along either $a$ or $b$ direction for the $ac$-plane (or $bc$-plane) full-breathing modes, as shown in Fig.~\ref{fig:demo}c. Figure~\ref{fig:vdWU} shows the comparison of the bare PBE and r2SCAN results and those from Hubbard U and D4 vdW corrections. Both bare PBE and r2SCAN results are too soft for the two challenging branches, and the PBE results are even softer, similar to the previous nonmagnetic results from r2SCAN\cite{YBCO6_NJL}. This is consistent with the fact that bare PBE cannot stabilize the AFM ground state and overestimates lattice constants. With U and vdW corrections, notable improvements are achieved for both PBE and r2SCAN, and the improvement is more significant for PBE than for r2SCAN, although the r2SCAN+U+D4 results remain closest to experiment. This is consistent with the observations and reasons discussed above for the structural and magnetic properties. To summarize, due to self-interaction reduction and capture of more intermediate range vdW interactions, r2SCAN performs much better than PBE in magnetic, structural and lattice dynamics properties of YBa$_2$Cu$_3$O$_6$, and for that reason the improvements from vdW and Hubbard U corrections are more significant for PBE than \btt{for} r2SCAN.

Figure~\ref{fig:TMs} includes the bare TMs and Hubbard U corrected results, compared with r2SCAN+U+D4 and experimental results. All the bare TMs results are too soft compared to those from r2SCAN, consistent with the fact that r2SCAN gives the best bare DFA predictions in basic magnetic and structural properties which is critical for accurate \btt{predictions of} lattice dynamics predictions. With Hubbard U correction, the improvement is significant and the results are close to but still softer than the r2SCAN+U+D4 results. Therefore r2SCAN generally has better performance than TMs for YBa$_2$Cu$_3$O$_6$ in structural, magnetic and lattice dynamics properties. This could be attributed to the more nonlocal nature of r2SCAN compared to TMs, since nonlocality is important for descriptions of semiconductors and insulators, \btt{as well as for more self-interaction reductions in DFA constructions}. Although both are at the same meta-GGA level, r2SCAN is more kinetic energy density-dependent and thus displays more nonlocality, while the TMs are more density gradient-dependent and less \btt{fully-nonlocal}. \btt{This argument \btt{for} the significance of nonlocality is consistent with an interesting and effective model proposed by Falter \textit{et al.}\cite{model_Falter1,model_Falter2,model_Falter3}, which separates the roles of local and nonlocal charge responses in phonons of cuprates.} 

\btt{\btt{Figure~\ref{fig:full}} shows the experimental and calculated phonon dispersions for all optical and acoustic modes. Note that there are soft acoustic modes for all the \btt{TM-family} functionals, with imaginary frequencies signaling structural instabilities predicted by \btt{them}.}

\FloatBarrier

\section{\label{sec:level1}Conclusions}
In summary, we have extended our previous work to a first-principles comparative study of the most challenging and experimentally important full-breathing modes of YBa$_2$Cu$_3$O$_6$. We achieve further understanding of both the lattice dynamics side and the density functional side. By applying both Hubbard U and the D4 vdW corrections to PBE and r2SCAN, notable improvements are obtained for structural, electronic, magnetic, and phonon dispersion predictions of YBa$_2$Cu$_3$O$_6$. The improvements from the combined corrections are more significant for PBE than for r2SCAN. With the improvements, PBE+U+D4 gives much better full-breathing phonon frequencies, closer but still softer compared to those from r2SCAN+U+D4 and experimental observations. Considering the general self-interaction reduction and \rtt{capture of more intermediate-range vdW in r2SCAN, in comparison with PBE,} we demonstrate the importance of vdW interactions and SIC in accurate YBa$_2$Cu$_3$O$_6$ lattice dynamics from first-principles, which in turn contributes to the major reason for the superior overall performance of r2SCAN over PBE. 

In addition, for the family of Tao-Mo meta-GGAs, all the bare DFA results are too soft compared to those from r2SCAN. With similar Hubbard U corrections, improvements are notable but still not enough to be as good as r2SCAN with corrections. Therefore r2SCAN generally has better performance than TMs for YBa$_2$Cu$_3$O$_6$ in structural, magnetic and lattice dynamics properties, which we attribute to the more nonlocal nature of r2SCAN compared to TMs. Nevertheless, the TMs could perform better\cite{Aaron_OFR2} for doped or gapless systems such as YBa$_2$Cu$_3$O$_7$\cite{YBCO7_anomaly,yubostripe,YBCO7LDA}, as implied by their good performances in surface, vacancy, and magnetic properties for metals\cite{TM_metals,revTM}, which could be further studied in the future. \rtt{Kaplan and Perdew\cite{Aaron_OFR2} argued that the perfect long-range screening of the exact exchange hole by the exact correlation hole in a metal can be  better captured by the semi-locality \btt{of} PBE (and by extension of the TM functionals) than by the fully-nonlocal density dependence of SCAN and r2SCAN (or the stronger full non-locality of many hybrid functionals).}

Note that, even with the best results we can achieve from r2SCAN+U+D4, there still exists noticeable softening for the tested full-breathing branches, especially for the peak at $\textbf{\textit{k}}\sim(0.2, 0, 0)$ along the experimental $ac$-plane full-breathing branch. \rtt{These residual discrepancies could imply extra physics or effects \btt{not included}, such as lattice anharmonicity, dynamical multiferroicity \cite{dyn_multiferro}, and hidden order beyond the simple antiferromagnetic ground state\cite{de2021,de2021b}.} Recent research efforts have renewed interest in the role of electron-phonon coupling in the mechanism of high-temperature superconductivity in cuprates \cite{Ahmadova2020}. The findings in the current work provide insights for future first-principles investigations on cuprates, including phonon anomalies\cite{bcsnotwork,giustino08nature,YBCO7LDA}, charge inhomogeneities, cavity-phonon-magnon quasiparticle interactions \cite{cavity-phonon-magnon}, and phase competition, which in turn contribute to a better understanding of high temperature superconducting materials.

\begin{acknowledgments}
J.N. and J.S. acknowledge the support of the U.S. Office of Naval Research (ONR) under Grant No. N00014-22-1-2673, with which they designed the project. J.N., A.R. and J.P.P acknowledge the support from Tulane University’s startup funds (computations). J.P.P. acknowledges the support from the National Science Foundation under Grant. No. DMR-1939528. The computational work done at Tulane University was supported by the Cypress Computational Cluster at Tulane and the National Energy Research Scientific Computing Center. \btt{The work at Northeastern University was supported by the US Department of Energy (DOE), Office of Science, Basic Energy Sciences Grant No. DE-SC0022216 (accurate modeling of complex materials) and benefited from Northeastern University’s Advanced Scientific Computation Center and the Discovery Cluster and the National Energy Research Scientific Computing Center through DOE Grant No. DE-AC02-05CH11231.} The work at Los Alamos National Laboratory was carried out under the auspices of the US Department of Energy (DOE) National Nuclear Security Administration under Contract No. 89233218CNA000001. It was supported by the LANL LDRD Program, the Quantum Science Center, a U.S. DOE Office of Science National Quantum Information Science Research Center, and in part by the Center for Integrated Nanotechnologies, a DOE BES user facility, in partnership with the LANL Institutional Computing Program for computational resources. B.B. was supported by the Ministry of Education and Culture (Finland) and by the LUT University INERCOM platform.
\end{acknowledgments}

\section*{Data Availability Statement}
The data that support the findings of this study are available from the corresponding author upon reasonable request.

\bibliography{ref}

\providecommand{\noopsort}[1]{}\providecommand{\singleletter}[1]{#1}%
\begin{thebibliography}{81}%
\makeatletter
\providecommand \@ifxundefined [1]{%
 \@ifx{#1\undefined}
}%
\providecommand \@ifnum [1]{%
 \ifnum #1\expandafter \@firstoftwo
 \else \expandafter \@secondoftwo
 \fi
}%
\providecommand \@ifx [1]{%
 \ifx #1\expandafter \@firstoftwo
 \else \expandafter \@secondoftwo
 \fi
}%
\providecommand \natexlab [1]{#1}%
\providecommand \enquote  [1]{``#1''}%
\providecommand \bibnamefont  [1]{#1}%
\providecommand \bibfnamefont [1]{#1}%
\providecommand \citenamefont [1]{#1}%
\providecommand \href@noop [0]{\@secondoftwo}%
\providecommand \href [0]{\begingroup \@sanitize@url \@href}%
\providecommand \@href[1]{\@@startlink{#1}\@@href}%
\providecommand \@@href[1]{\endgroup#1\@@endlink}%
\providecommand \@sanitize@url [0]{\catcode `\\12\catcode `\$12\catcode `\&12\catcode `\#12\catcode `\^12\catcode `\_12\catcode `\%12\relax}%
\providecommand \@@startlink[1]{}%
\providecommand \@@endlink[0]{}%
\providecommand \url  [0]{\begingroup\@sanitize@url \@url }%
\providecommand \@url [1]{\endgroup\@href {#1}{\urlprefix }}%
\providecommand \urlprefix  [0]{URL }%
\providecommand \Eprint [0]{\href }%
\providecommand \doibase [0]{http://dx.doi.org/}%
\providecommand \selectlanguage [0]{\@gobble}%
\providecommand \bibinfo  [0]{\@secondoftwo}%
\providecommand \bibfield  [0]{\@secondoftwo}%
\providecommand \translation [1]{[#1]}%
\providecommand \BibitemOpen [0]{}%
\providecommand \bibitemStop [0]{}%
\providecommand \bibitemNoStop [0]{.\EOS\space}%
\providecommand \EOS [0]{\spacefactor3000\relax}%
\providecommand \BibitemShut  [1]{\csname bibitem#1\endcsname}%
\let\auto@bib@innerbib\@empty
\bibitem [{\citenamefont {Bednorz}\ and\ \citenamefont {M{\"u}ller}(1986)}]{1986HTc}%
  \BibitemOpen
  \bibfield  {author} {\bibinfo {author} {\bibfnamefont {J.~G.}\ \bibnamefont {Bednorz}}\ and\ \bibinfo {author} {\bibfnamefont {K.~A.}\ \bibnamefont {M{\"u}ller}},\ }\bibfield  {title} {\enquote {\bibinfo {title} {{Possible high T$_{C}$ superconductivity in the Ba-La-Cu-O system}},}\ }\href@noop {} {\bibfield  {journal} {\bibinfo  {journal} {Zeitschrift f{\"u}r Physik B Condensed Matter}\ }\textbf {\bibinfo {volume} {64}},\ \bibinfo {pages} {189--193} (\bibinfo {year} {1986})}\BibitemShut {NoStop}%
\bibitem [{\citenamefont {Anderson}(1987)}]{RVBAnderson}%
  \BibitemOpen
  \bibfield  {author} {\bibinfo {author} {\bibfnamefont {P.~W.}\ \bibnamefont {Anderson}},\ }\bibfield  {title} {\enquote {\bibinfo {title} {{The resonating valence bond state in La$_{2}$CuO$_{4}$ and superconductivity}},}\ }\href@noop {} {\bibfield  {journal} {\bibinfo  {journal} {Science}\ }\textbf {\bibinfo {volume} {235}},\ \bibinfo {pages} {1196--1198} (\bibinfo {year} {1987})}\BibitemShut {NoStop}%
\bibitem [{\citenamefont {Tsuei}\ and\ \citenamefont {Kirtley}(2000)}]{pairRMP}%
  \BibitemOpen
  \bibfield  {author} {\bibinfo {author} {\bibfnamefont {C.~C.}\ \bibnamefont {Tsuei}}\ and\ \bibinfo {author} {\bibfnamefont {J.~R.}\ \bibnamefont {Kirtley}},\ }\bibfield  {title} {\enquote {\bibinfo {title} {Pairing symmetry in cuprate superconductors},}\ }\href {\doibase 10.1103/RevModPhys.72.969} {\bibfield  {journal} {\bibinfo  {journal} {Rev. Mod. Phys.}\ }\textbf {\bibinfo {volume} {72}},\ \bibinfo {pages} {969--1016} (\bibinfo {year} {2000})}\BibitemShut {NoStop}%
\bibitem [{\citenamefont {Damascelli}, \citenamefont {Hussain},\ and\ \citenamefont {Shen}(2003)}]{arpsRMP}%
  \BibitemOpen
  \bibfield  {author} {\bibinfo {author} {\bibfnamefont {A.}~\bibnamefont {Damascelli}}, \bibinfo {author} {\bibfnamefont {Z.}~\bibnamefont {Hussain}}, \ and\ \bibinfo {author} {\bibfnamefont {Z.-X.}\ \bibnamefont {Shen}},\ }\bibfield  {title} {\enquote {\bibinfo {title} {Angle-resolved photoemission studies of the cuprate superconductors},}\ }\href {\doibase 10.1103/RevModPhys.75.473} {\bibfield  {journal} {\bibinfo  {journal} {Rev. Mod. Phys.}\ }\textbf {\bibinfo {volume} {75}},\ \bibinfo {pages} {473--541} (\bibinfo {year} {2003})}\BibitemShut {NoStop}%
\bibitem [{\citenamefont {Keimer}\ \emph {et~al.}(2015)\citenamefont {Keimer}, \citenamefont {Kivelson}, \citenamefont {Norman}, \citenamefont {Uchida},\ and\ \citenamefont {Zaanen}}]{keimer2015}%
  \BibitemOpen
  \bibfield  {author} {\bibinfo {author} {\bibfnamefont {B.}~\bibnamefont {Keimer}}, \bibinfo {author} {\bibfnamefont {S.~A.}\ \bibnamefont {Kivelson}}, \bibinfo {author} {\bibfnamefont {M.~R.}\ \bibnamefont {Norman}}, \bibinfo {author} {\bibfnamefont {S.}~\bibnamefont {Uchida}}, \ and\ \bibinfo {author} {\bibfnamefont {J.}~\bibnamefont {Zaanen}},\ }\bibfield  {title} {\enquote {\bibinfo {title} {From quantum matter to high-temperature superconductivity in copper oxides},}\ }\href@noop {} {\bibfield  {journal} {\bibinfo  {journal} {Nature}\ }\textbf {\bibinfo {volume} {518}},\ \bibinfo {pages} {179--186} (\bibinfo {year} {2015})}\BibitemShut {NoStop}%
\bibitem [{\citenamefont {Sobota}, \citenamefont {He},\ and\ \citenamefont {Shen}(2021)}]{Sobota2021}%
  \BibitemOpen
  \bibfield  {author} {\bibinfo {author} {\bibfnamefont {J.~A.}\ \bibnamefont {Sobota}}, \bibinfo {author} {\bibfnamefont {Y.}~\bibnamefont {He}}, \ and\ \bibinfo {author} {\bibfnamefont {Z.-X.}\ \bibnamefont {Shen}},\ }\bibfield  {title} {\enquote {\bibinfo {title} {Angle-resolved photoemission studies of quantum materials},}\ }\href {\doibase 10.1103/RevModPhys.93.025006} {\bibfield  {journal} {\bibinfo  {journal} {Rev. Mod. Phys.}\ }\textbf {\bibinfo {volume} {93}},\ \bibinfo {pages} {025006} (\bibinfo {year} {2021})}\BibitemShut {NoStop}%
\bibitem [{\citenamefont {Bohnen}, \citenamefont {Heid},\ and\ \citenamefont {Krauss}(2003)}]{bcsnotwork}%
  \BibitemOpen
  \bibfield  {author} {\bibinfo {author} {\bibfnamefont {K.-P.}\ \bibnamefont {Bohnen}}, \bibinfo {author} {\bibfnamefont {R.}~\bibnamefont {Heid}}, \ and\ \bibinfo {author} {\bibfnamefont {M.}~\bibnamefont {Krauss}},\ }\bibfield  {title} {\enquote {\bibinfo {title} {{Phonon dispersion and electron-phonon interaction for YBa$_2$Cu$_3$O$_7$ from first-principles calculations}},}\ }\href@noop {} {\bibfield  {journal} {\bibinfo  {journal} {EPL (Europhysics Letters)}\ }\textbf {\bibinfo {volume} {64}},\ \bibinfo {pages} {104} (\bibinfo {year} {2003})}\BibitemShut {NoStop}%
\bibitem [{\citenamefont {Giustino}, \citenamefont {Cohen},\ and\ \citenamefont {Louie}(2008)}]{giustino08nature}%
  \BibitemOpen
  \bibfield  {author} {\bibinfo {author} {\bibfnamefont {F.}~\bibnamefont {Giustino}}, \bibinfo {author} {\bibfnamefont {M.~L.}\ \bibnamefont {Cohen}}, \ and\ \bibinfo {author} {\bibfnamefont {S.~G.}\ \bibnamefont {Louie}},\ }\bibfield  {title} {\enquote {\bibinfo {title} {Small phonon contribution to the photoemission kink in the copper oxide superconductors},}\ }\href@noop {} {\bibfield  {journal} {\bibinfo  {journal} {Nature}\ }\textbf {\bibinfo {volume} {452}},\ \bibinfo {pages} {975--978} (\bibinfo {year} {2008})}\BibitemShut {NoStop}%
\bibitem [{\citenamefont {Heid}\ \emph {et~al.}(2009)\citenamefont {Heid}, \citenamefont {Zeyher}, \citenamefont {Manske},\ and\ \citenamefont {Bohnen}}]{YBCO7LDA}%
  \BibitemOpen
  \bibfield  {author} {\bibinfo {author} {\bibfnamefont {R.}~\bibnamefont {Heid}}, \bibinfo {author} {\bibfnamefont {R.}~\bibnamefont {Zeyher}}, \bibinfo {author} {\bibfnamefont {D.}~\bibnamefont {Manske}}, \ and\ \bibinfo {author} {\bibfnamefont {K.-P.}\ \bibnamefont {Bohnen}},\ }\bibfield  {title} {\enquote {\bibinfo {title} {{Phonon-induced pairing interaction in ${\text{YBa}}_{2}{\text{Cu}}_{3}{\text{O}}_{7}$ within the local-density approximation}},}\ }\href {\doibase 10.1103/PhysRevB.80.024507} {\bibfield  {journal} {\bibinfo  {journal} {Phys. Rev. B}\ }\textbf {\bibinfo {volume} {80}},\ \bibinfo {pages} {024507} (\bibinfo {year} {2009})}\BibitemShut {NoStop}%
\bibitem [{\citenamefont {Cooper}(1956)}]{BCS1}%
  \BibitemOpen
  \bibfield  {author} {\bibinfo {author} {\bibfnamefont {L.~N.}\ \bibnamefont {Cooper}},\ }\bibfield  {title} {\enquote {\bibinfo {title} {Bound electron pairs in a degenerate fermi gas},}\ }\href {\doibase 10.1103/PhysRev.104.1189} {\bibfield  {journal} {\bibinfo  {journal} {Phys. Rev.}\ }\textbf {\bibinfo {volume} {104}},\ \bibinfo {pages} {1189--1190} (\bibinfo {year} {1956})}\BibitemShut {NoStop}%
\bibitem [{\citenamefont {Bardeen}, \citenamefont {Cooper},\ and\ \citenamefont {Schrieffer}(1957{\natexlab{a}})}]{BCS2}%
  \BibitemOpen
  \bibfield  {author} {\bibinfo {author} {\bibfnamefont {J.}~\bibnamefont {Bardeen}}, \bibinfo {author} {\bibfnamefont {L.~N.}\ \bibnamefont {Cooper}}, \ and\ \bibinfo {author} {\bibfnamefont {J.~R.}\ \bibnamefont {Schrieffer}},\ }\bibfield  {title} {\enquote {\bibinfo {title} {Microscopic theory of superconductivity},}\ }\href {\doibase 10.1103/PhysRev.106.162} {\bibfield  {journal} {\bibinfo  {journal} {Phys. Rev.}\ }\textbf {\bibinfo {volume} {106}},\ \bibinfo {pages} {162--164} (\bibinfo {year} {1957}{\natexlab{a}})}\BibitemShut {NoStop}%
\bibitem [{\citenamefont {Bardeen}, \citenamefont {Cooper},\ and\ \citenamefont {Schrieffer}(1957{\natexlab{b}})}]{BCS3}%
  \BibitemOpen
  \bibfield  {author} {\bibinfo {author} {\bibfnamefont {J.}~\bibnamefont {Bardeen}}, \bibinfo {author} {\bibfnamefont {L.~N.}\ \bibnamefont {Cooper}}, \ and\ \bibinfo {author} {\bibfnamefont {J.~R.}\ \bibnamefont {Schrieffer}},\ }\bibfield  {title} {\enquote {\bibinfo {title} {Theory of superconductivity},}\ }\href {\doibase 10.1103/PhysRev.108.1175} {\bibfield  {journal} {\bibinfo  {journal} {Phys. Rev.}\ }\textbf {\bibinfo {volume} {108}},\ \bibinfo {pages} {1175--1204} (\bibinfo {year} {1957}{\natexlab{b}})}\BibitemShut {NoStop}%
\bibitem [{\citenamefont {McQueeney}\ \emph {et~al.}(2001)\citenamefont {McQueeney}, \citenamefont {Sarrao}, \citenamefont {Pagliuso}, \citenamefont {Stephens},\ and\ \citenamefont {Osborn}}]{ELmixHTc}%
  \BibitemOpen
  \bibfield  {author} {\bibinfo {author} {\bibfnamefont {R.~J.}\ \bibnamefont {McQueeney}}, \bibinfo {author} {\bibfnamefont {J.~L.}\ \bibnamefont {Sarrao}}, \bibinfo {author} {\bibfnamefont {P.~G.}\ \bibnamefont {Pagliuso}}, \bibinfo {author} {\bibfnamefont {P.~W.}\ \bibnamefont {Stephens}}, \ and\ \bibinfo {author} {\bibfnamefont {R.}~\bibnamefont {Osborn}},\ }\bibfield  {title} {\enquote {\bibinfo {title} {Mixed lattice and electronic states in high-temperature superconductors},}\ }\href {\doibase 10.1103/PhysRevLett.87.077001} {\bibfield  {journal} {\bibinfo  {journal} {Phys. Rev. Lett.}\ }\textbf {\bibinfo {volume} {87}},\ \bibinfo {pages} {077001} (\bibinfo {year} {2001})}\BibitemShut {NoStop}%
\bibitem [{\citenamefont {Reznik}\ \emph {et~al.}(2006)\citenamefont {Reznik}, \citenamefont {Pintschovius}, \citenamefont {Ito}, \citenamefont {Iikubo}, \citenamefont {Sato}, \citenamefont {Goka}, \citenamefont {Fujita}, \citenamefont {Yamada}, \citenamefont {Gu},\ and\ \citenamefont {Tranquada}}]{rezniknatureEPC}%
  \BibitemOpen
  \bibfield  {author} {\bibinfo {author} {\bibfnamefont {D.}~\bibnamefont {Reznik}}, \bibinfo {author} {\bibfnamefont {L.}~\bibnamefont {Pintschovius}}, \bibinfo {author} {\bibfnamefont {M.}~\bibnamefont {Ito}}, \bibinfo {author} {\bibfnamefont {S.}~\bibnamefont {Iikubo}}, \bibinfo {author} {\bibfnamefont {M.}~\bibnamefont {Sato}}, \bibinfo {author} {\bibfnamefont {H.}~\bibnamefont {Goka}}, \bibinfo {author} {\bibfnamefont {M.}~\bibnamefont {Fujita}}, \bibinfo {author} {\bibfnamefont {K.}~\bibnamefont {Yamada}}, \bibinfo {author} {\bibfnamefont {G.}~\bibnamefont {Gu}}, \ and\ \bibinfo {author} {\bibfnamefont {J.}~\bibnamefont {Tranquada}},\ }\bibfield  {title} {\enquote {\bibinfo {title} {Electron-phonon coupling reflecting dynamic charge inhomogeneity in copper oxide superconductors},}\ }\href@noop {} {\bibfield  {journal} {\bibinfo  {journal} {Nature}\ }\textbf {\bibinfo {volume} {440}},\ \bibinfo {pages} {1170--1173} (\bibinfo {year} {2006})}\BibitemShut {NoStop}%
\bibitem [{\citenamefont {Lanzara}\ \emph {et~al.}(2001)\citenamefont {Lanzara}, \citenamefont {Bogdanov}, \citenamefont {Zhou}, \citenamefont {Kellar}, \citenamefont {Feng}, \citenamefont {Lu}, \citenamefont {Yoshida}, \citenamefont {Eisaki}, \citenamefont {Fujimori}, \citenamefont {Kishio} \emph {et~al.}}]{natureEPC}%
  \BibitemOpen
  \bibfield  {author} {\bibinfo {author} {\bibfnamefont {A.}~\bibnamefont {Lanzara}}, \bibinfo {author} {\bibfnamefont {P.}~\bibnamefont {Bogdanov}}, \bibinfo {author} {\bibfnamefont {X.}~\bibnamefont {Zhou}}, \bibinfo {author} {\bibfnamefont {S.}~\bibnamefont {Kellar}}, \bibinfo {author} {\bibfnamefont {D.}~\bibnamefont {Feng}}, \bibinfo {author} {\bibfnamefont {E.}~\bibnamefont {Lu}}, \bibinfo {author} {\bibfnamefont {T.}~\bibnamefont {Yoshida}}, \bibinfo {author} {\bibfnamefont {H.}~\bibnamefont {Eisaki}}, \bibinfo {author} {\bibfnamefont {A.}~\bibnamefont {Fujimori}}, \bibinfo {author} {\bibfnamefont {K.}~\bibnamefont {Kishio}},  \emph {et~al.},\ }\bibfield  {title} {\enquote {\bibinfo {title} {Evidence for ubiquitous strong electron-phonon coupling in high-temperature superconductors},}\ }\href@noop {} {\bibfield  {journal} {\bibinfo  {journal} {Nature}\ }\textbf {\bibinfo {volume} {412}},\ \bibinfo {pages} {510--514} (\bibinfo {year} {2001})}\BibitemShut {NoStop}%
\bibitem [{\citenamefont {d'Astuto}\ \emph {et~al.}(2002)\citenamefont {d'Astuto}, \citenamefont {Mang}, \citenamefont {Giura}, \citenamefont {Shukla}, \citenamefont {Ghigna}, \citenamefont {Mirone}, \citenamefont {Braden}, \citenamefont {Greven}, \citenamefont {Krisch},\ and\ \citenamefont {Sette}}]{NCCOanomaly}%
  \BibitemOpen
  \bibfield  {author} {\bibinfo {author} {\bibfnamefont {M.}~\bibnamefont {d'Astuto}}, \bibinfo {author} {\bibfnamefont {P.~K.}\ \bibnamefont {Mang}}, \bibinfo {author} {\bibfnamefont {P.}~\bibnamefont {Giura}}, \bibinfo {author} {\bibfnamefont {A.}~\bibnamefont {Shukla}}, \bibinfo {author} {\bibfnamefont {P.}~\bibnamefont {Ghigna}}, \bibinfo {author} {\bibfnamefont {A.}~\bibnamefont {Mirone}}, \bibinfo {author} {\bibfnamefont {M.}~\bibnamefont {Braden}}, \bibinfo {author} {\bibfnamefont {M.}~\bibnamefont {Greven}}, \bibinfo {author} {\bibfnamefont {M.}~\bibnamefont {Krisch}}, \ and\ \bibinfo {author} {\bibfnamefont {F.}~\bibnamefont {Sette}},\ }\bibfield  {title} {\enquote {\bibinfo {title} {Anomalous dispersion of longitudinal optical phonons in {${\mathrm{Nd}}_{1.86}{\mathrm{Ce}}_{0.14}{\mathrm{CuO}}_{4+\ensuremath{\delta}}$} determined by inelastic x-ray scattering},}\ }\href {\doibase 10.1103/PhysRevLett.88.167002} {\bibfield  {journal} {\bibinfo  {journal} {Phys. Rev. Lett.}\ }\textbf {\bibinfo {volume}
  {88}},\ \bibinfo {pages} {167002} (\bibinfo {year} {2002})}\BibitemShut {NoStop}%
\bibitem [{\citenamefont {McQueeney}\ \emph {et~al.}(1999)\citenamefont {McQueeney}, \citenamefont {Petrov}, \citenamefont {Egami}, \citenamefont {Yethiraj}, \citenamefont {Shirane},\ and\ \citenamefont {Endoh}}]{LSCOanomaly}%
  \BibitemOpen
  \bibfield  {author} {\bibinfo {author} {\bibfnamefont {R.~J.}\ \bibnamefont {McQueeney}}, \bibinfo {author} {\bibfnamefont {Y.}~\bibnamefont {Petrov}}, \bibinfo {author} {\bibfnamefont {T.}~\bibnamefont {Egami}}, \bibinfo {author} {\bibfnamefont {M.}~\bibnamefont {Yethiraj}}, \bibinfo {author} {\bibfnamefont {G.}~\bibnamefont {Shirane}}, \ and\ \bibinfo {author} {\bibfnamefont {Y.}~\bibnamefont {Endoh}},\ }\bibfield  {title} {\enquote {\bibinfo {title} {{Anomalous dispersion of LO phonons in ${\mathrm{La}}_{1.85}{\mathrm{Sr}}_{0.15}{\mathrm{CuO}}_{4}$ at low temperatures}},}\ }\href {\doibase 10.1103/PhysRevLett.82.628} {\bibfield  {journal} {\bibinfo  {journal} {Phys. Rev. Lett.}\ }\textbf {\bibinfo {volume} {82}},\ \bibinfo {pages} {628--631} (\bibinfo {year} {1999})}\BibitemShut {NoStop}%
\bibitem [{\citenamefont {Gweon}\ \emph {et~al.}(2004)\citenamefont {Gweon}, \citenamefont {Sasagawa}, \citenamefont {Zhou}, \citenamefont {Graf}, \citenamefont {Takagi}, \citenamefont {Lee},\ and\ \citenamefont {Lanzara}}]{isotopeHTc}%
  \BibitemOpen
  \bibfield  {author} {\bibinfo {author} {\bibfnamefont {G.-H.}\ \bibnamefont {Gweon}}, \bibinfo {author} {\bibfnamefont {T.}~\bibnamefont {Sasagawa}}, \bibinfo {author} {\bibfnamefont {S.}~\bibnamefont {Zhou}}, \bibinfo {author} {\bibfnamefont {J.}~\bibnamefont {Graf}}, \bibinfo {author} {\bibfnamefont {H.}~\bibnamefont {Takagi}}, \bibinfo {author} {\bibfnamefont {D.-H.}\ \bibnamefont {Lee}}, \ and\ \bibinfo {author} {\bibfnamefont {A.}~\bibnamefont {Lanzara}},\ }\bibfield  {title} {\enquote {\bibinfo {title} {An unusual isotope effect in a high-transition-temperature superconductor},}\ }\href@noop {} {\bibfield  {journal} {\bibinfo  {journal} {Nature}\ }\textbf {\bibinfo {volume} {430}},\ \bibinfo {pages} {187--190} (\bibinfo {year} {2004})}\BibitemShut {NoStop}%
\bibitem [{\citenamefont {He}\ \emph {et~al.}(2018)\citenamefont {He}, \citenamefont {Hashimoto}, \citenamefont {Song}, \citenamefont {Chen}, \citenamefont {He}, \citenamefont {Vishik}, \citenamefont {Moritz}, \citenamefont {Lee}, \citenamefont {Nagaosa}, \citenamefont {Zaanen} \emph {et~al.}}]{B2212EPC}%
  \BibitemOpen
  \bibfield  {author} {\bibinfo {author} {\bibfnamefont {Y.}~\bibnamefont {He}}, \bibinfo {author} {\bibfnamefont {M.}~\bibnamefont {Hashimoto}}, \bibinfo {author} {\bibfnamefont {D.}~\bibnamefont {Song}}, \bibinfo {author} {\bibfnamefont {S.-D.}\ \bibnamefont {Chen}}, \bibinfo {author} {\bibfnamefont {J.}~\bibnamefont {He}}, \bibinfo {author} {\bibfnamefont {I.}~\bibnamefont {Vishik}}, \bibinfo {author} {\bibfnamefont {B.}~\bibnamefont {Moritz}}, \bibinfo {author} {\bibfnamefont {D.-H.}\ \bibnamefont {Lee}}, \bibinfo {author} {\bibfnamefont {N.}~\bibnamefont {Nagaosa}}, \bibinfo {author} {\bibfnamefont {J.}~\bibnamefont {Zaanen}},  \emph {et~al.},\ }\bibfield  {title} {\enquote {\bibinfo {title} {{Rapid change of superconductivity and electron-phonon coupling through critical doping in Bi-2212}},}\ }\href@noop {} {\bibfield  {journal} {\bibinfo  {journal} {Science}\ }\textbf {\bibinfo {volume} {362}},\ \bibinfo {pages} {62--65} (\bibinfo {year} {2018})}\BibitemShut {NoStop}%
\bibitem [{\citenamefont {Dal~Conte}\ \emph {et~al.}(2012)\citenamefont {Dal~Conte}, \citenamefont {Giannetti}, \citenamefont {Coslovich}, \citenamefont {Cilento}, \citenamefont {Bossini}, \citenamefont {Abebaw}, \citenamefont {Banfi}, \citenamefont {Ferrini}, \citenamefont {Eisaki}, \citenamefont {Greven} \emph {et~al.}}]{dal2012}%
  \BibitemOpen
  \bibfield  {author} {\bibinfo {author} {\bibfnamefont {S.}~\bibnamefont {Dal~Conte}}, \bibinfo {author} {\bibfnamefont {C.}~\bibnamefont {Giannetti}}, \bibinfo {author} {\bibfnamefont {G.}~\bibnamefont {Coslovich}}, \bibinfo {author} {\bibfnamefont {F.}~\bibnamefont {Cilento}}, \bibinfo {author} {\bibfnamefont {D.}~\bibnamefont {Bossini}}, \bibinfo {author} {\bibfnamefont {T.}~\bibnamefont {Abebaw}}, \bibinfo {author} {\bibfnamefont {F.}~\bibnamefont {Banfi}}, \bibinfo {author} {\bibfnamefont {G.}~\bibnamefont {Ferrini}}, \bibinfo {author} {\bibfnamefont {H.}~\bibnamefont {Eisaki}}, \bibinfo {author} {\bibfnamefont {M.}~\bibnamefont {Greven}},  \emph {et~al.},\ }\bibfield  {title} {\enquote {\bibinfo {title} {{Disentangling the electronic and phononic glue in a high-T$_{C}$ superconductor}},}\ }\href@noop {} {\bibfield  {journal} {\bibinfo  {journal} {Science}\ }\textbf {\bibinfo {volume} {335}},\ \bibinfo {pages} {1600--1603} (\bibinfo {year} {2012})}\BibitemShut {NoStop}%
\bibitem [{\citenamefont {Zhang}\ \emph {et~al.}(2020{\natexlab{a}})\citenamefont {Zhang}, \citenamefont {Lane}, \citenamefont {Furness}, \citenamefont {Barbiellini}, \citenamefont {Perdew}, \citenamefont {Markiewicz}, \citenamefont {Bansil},\ and\ \citenamefont {Sun}}]{yubostripe}%
  \BibitemOpen
  \bibfield  {author} {\bibinfo {author} {\bibfnamefont {Y.}~\bibnamefont {Zhang}}, \bibinfo {author} {\bibfnamefont {C.}~\bibnamefont {Lane}}, \bibinfo {author} {\bibfnamefont {J.~W.}\ \bibnamefont {Furness}}, \bibinfo {author} {\bibfnamefont {B.}~\bibnamefont {Barbiellini}}, \bibinfo {author} {\bibfnamefont {J.~P.}\ \bibnamefont {Perdew}}, \bibinfo {author} {\bibfnamefont {R.~S.}\ \bibnamefont {Markiewicz}}, \bibinfo {author} {\bibfnamefont {A.}~\bibnamefont {Bansil}}, \ and\ \bibinfo {author} {\bibfnamefont {J.}~\bibnamefont {Sun}},\ }\bibfield  {title} {\enquote {\bibinfo {title} {Competing stripe and magnetic phases in the cuprates from first principles},}\ }\href@noop {} {\bibfield  {journal} {\bibinfo  {journal} {Proceedings of the National Academy of Sciences}\ }\textbf {\bibinfo {volume} {117}},\ \bibinfo {pages} {68--72} (\bibinfo {year} {2020}{\natexlab{a}})}\BibitemShut {NoStop}%
\bibitem [{\citenamefont {Furness}\ \emph {et~al.}(2018)\citenamefont {Furness}, \citenamefont {Zhang}, \citenamefont {Lane}, \citenamefont {Buda}, \citenamefont {Barbiellini}, \citenamefont {Markiewicz}, \citenamefont {Bansil},\ and\ \citenamefont {Sun}}]{James_SCAN_Cuprates}%
  \BibitemOpen
  \bibfield  {author} {\bibinfo {author} {\bibfnamefont {J.~W.}\ \bibnamefont {Furness}}, \bibinfo {author} {\bibfnamefont {Y.}~\bibnamefont {Zhang}}, \bibinfo {author} {\bibfnamefont {C.}~\bibnamefont {Lane}}, \bibinfo {author} {\bibfnamefont {I.~G.}\ \bibnamefont {Buda}}, \bibinfo {author} {\bibfnamefont {B.}~\bibnamefont {Barbiellini}}, \bibinfo {author} {\bibfnamefont {R.~S.}\ \bibnamefont {Markiewicz}}, \bibinfo {author} {\bibfnamefont {A.}~\bibnamefont {Bansil}}, \ and\ \bibinfo {author} {\bibfnamefont {J.}~\bibnamefont {Sun}},\ }\bibfield  {title} {\enquote {\bibinfo {title} {An accurate first-principles treatment of doping-dependent electronic structure of high-temperature cuprate superconductors},}\ }\href@noop {} {\bibfield  {journal} {\bibinfo  {journal} {Communications Physics}\ }\textbf {\bibinfo {volume} {1}},\ \bibinfo {pages} {1--6} (\bibinfo {year} {2018})}\BibitemShut {NoStop}%
\bibitem [{\citenamefont {Lane}\ \emph {et~al.}(2018)\citenamefont {Lane}, \citenamefont {Furness}, \citenamefont {Buda}, \citenamefont {Zhang}, \citenamefont {Markiewicz}, \citenamefont {Barbiellini}, \citenamefont {Sun},\ and\ \citenamefont {Bansil}}]{Chris_LCO}%
  \BibitemOpen
  \bibfield  {author} {\bibinfo {author} {\bibfnamefont {C.}~\bibnamefont {Lane}}, \bibinfo {author} {\bibfnamefont {J.~W.}\ \bibnamefont {Furness}}, \bibinfo {author} {\bibfnamefont {I.~G.}\ \bibnamefont {Buda}}, \bibinfo {author} {\bibfnamefont {Y.}~\bibnamefont {Zhang}}, \bibinfo {author} {\bibfnamefont {R.~S.}\ \bibnamefont {Markiewicz}}, \bibinfo {author} {\bibfnamefont {B.}~\bibnamefont {Barbiellini}}, \bibinfo {author} {\bibfnamefont {J.}~\bibnamefont {Sun}}, \ and\ \bibinfo {author} {\bibfnamefont {A.}~\bibnamefont {Bansil}},\ }\bibfield  {title} {\enquote {\bibinfo {title} {Antiferromagnetic ground state of {${\mathrm{La}}_{2}{\mathrm{CuO}}_{4}$}: A parameter-free ab initio description},}\ }\href {\doibase 10.1103/PhysRevB.98.125140} {\bibfield  {journal} {\bibinfo  {journal} {Phys. Rev. B}\ }\textbf {\bibinfo {volume} {98}},\ \bibinfo {pages} {125140} (\bibinfo {year} {2018})}\BibitemShut {NoStop}%
\bibitem [{\citenamefont {Anisimov}, \citenamefont {Zaanen},\ and\ \citenamefont {Andersen}(1991)}]{HubbardU}%
  \BibitemOpen
  \bibfield  {author} {\bibinfo {author} {\bibfnamefont {V.~I.}\ \bibnamefont {Anisimov}}, \bibinfo {author} {\bibfnamefont {J.}~\bibnamefont {Zaanen}}, \ and\ \bibinfo {author} {\bibfnamefont {O.~K.}\ \bibnamefont {Andersen}},\ }\bibfield  {title} {\enquote {\bibinfo {title} {{Band theory and Mott insulators: Hubbard U instead of Stoner I}},}\ }\href {\doibase 10.1103/PhysRevB.44.943} {\bibfield  {journal} {\bibinfo  {journal} {Phys. Rev. B}\ }\textbf {\bibinfo {volume} {44}},\ \bibinfo {pages} {943--954} (\bibinfo {year} {1991})}\BibitemShut {NoStop}%
\bibitem [{\citenamefont {Cococcioni}\ and\ \citenamefont {de~Gironcoli}(2005)}]{linearU}%
  \BibitemOpen
  \bibfield  {author} {\bibinfo {author} {\bibfnamefont {M.}~\bibnamefont {Cococcioni}}\ and\ \bibinfo {author} {\bibfnamefont {S.}~\bibnamefont {de~Gironcoli}},\ }\bibfield  {title} {\enquote {\bibinfo {title} {Linear response approach to the calculation of the effective interaction parameters in the $\mathrm{LDA}+\mathrm{U}$ method},}\ }\href {\doibase 10.1103/PhysRevB.71.035105} {\bibfield  {journal} {\bibinfo  {journal} {Phys. Rev. B}\ }\textbf {\bibinfo {volume} {71}},\ \bibinfo {pages} {035105} (\bibinfo {year} {2005})}\BibitemShut {NoStop}%
\bibitem [{\citenamefont {Kasinathan}\ \emph {et~al.}(2006)\citenamefont {Kasinathan}, \citenamefont {Kune\ifmmode~\check{s}\else \v{s}\fi{}}, \citenamefont {Koepernik}, \citenamefont {Diaconu}, \citenamefont {Martin}, \citenamefont {Prodan}, \citenamefont {Scuseria}, \citenamefont {Spaldin}, \citenamefont {Petit}, \citenamefont {Schulthess},\ and\ \citenamefont {Pickett}}]{LDAU_MnO}%
  \BibitemOpen
  \bibfield  {author} {\bibinfo {author} {\bibfnamefont {D.}~\bibnamefont {Kasinathan}}, \bibinfo {author} {\bibfnamefont {J.}~\bibnamefont {Kune\ifmmode~\check{s}\else \v{s}\fi{}}}, \bibinfo {author} {\bibfnamefont {K.}~\bibnamefont {Koepernik}}, \bibinfo {author} {\bibfnamefont {C.~V.}\ \bibnamefont {Diaconu}}, \bibinfo {author} {\bibfnamefont {R.~L.}\ \bibnamefont {Martin}}, \bibinfo {author} {\bibfnamefont {I.~m. c.~D.}\ \bibnamefont {Prodan}}, \bibinfo {author} {\bibfnamefont {G.~E.}\ \bibnamefont {Scuseria}}, \bibinfo {author} {\bibfnamefont {N.}~\bibnamefont {Spaldin}}, \bibinfo {author} {\bibfnamefont {L.}~\bibnamefont {Petit}}, \bibinfo {author} {\bibfnamefont {T.~C.}\ \bibnamefont {Schulthess}}, \ and\ \bibinfo {author} {\bibfnamefont {W.~E.}\ \bibnamefont {Pickett}},\ }\bibfield  {title} {\enquote {\bibinfo {title} {Mott transition of mno under pressure: A comparison of correlated band theories},}\ }\href {\doibase 10.1103/PhysRevB.74.195110} {\bibfield  {journal} {\bibinfo  {journal} {Phys. Rev.
  B}\ }\textbf {\bibinfo {volume} {74}},\ \bibinfo {pages} {195110} (\bibinfo {year} {2006})}\BibitemShut {NoStop}%
\bibitem [{\citenamefont {Dudarev}\ \emph {et~al.}(1998)\citenamefont {Dudarev}, \citenamefont {Botton}, \citenamefont {Savrasov}, \citenamefont {Humphreys},\ and\ \citenamefont {Sutton}}]{LSDAU_NiO}%
  \BibitemOpen
  \bibfield  {author} {\bibinfo {author} {\bibfnamefont {S.~L.}\ \bibnamefont {Dudarev}}, \bibinfo {author} {\bibfnamefont {G.~A.}\ \bibnamefont {Botton}}, \bibinfo {author} {\bibfnamefont {S.~Y.}\ \bibnamefont {Savrasov}}, \bibinfo {author} {\bibfnamefont {C.~J.}\ \bibnamefont {Humphreys}}, \ and\ \bibinfo {author} {\bibfnamefont {A.~P.}\ \bibnamefont {Sutton}},\ }\bibfield  {title} {\enquote {\bibinfo {title} {{Electron-energy-loss spectra and the structural stability of nickel oxide: An LSDA+U study}},}\ }\href {\doibase 10.1103/PhysRevB.57.1505} {\bibfield  {journal} {\bibinfo  {journal} {Phys. Rev. B}\ }\textbf {\bibinfo {volume} {57}},\ \bibinfo {pages} {1505--1509} (\bibinfo {year} {1998})}\BibitemShut {NoStop}%
\bibitem [{\citenamefont {Sterling}\ and\ \citenamefont {Reznik}(2021)}]{sterling2021}%
  \BibitemOpen
  \bibfield  {author} {\bibinfo {author} {\bibfnamefont {T.~C.}\ \bibnamefont {Sterling}}\ and\ \bibinfo {author} {\bibfnamefont {D.}~\bibnamefont {Reznik}},\ }\bibfield  {title} {\enquote {\bibinfo {title} {{Effect of the electronic charge gap on LO bond-stretching phonons in undoped La$_2$CuO$_4$ calculated using LDA+U}},}\ }\href@noop {} {\bibfield  {journal} {\bibinfo  {journal} {Physical Review B}\ }\textbf {\bibinfo {volume} {104}},\ \bibinfo {pages} {134311} (\bibinfo {year} {2021})}\BibitemShut {NoStop}%
\bibitem [{\citenamefont {Jarlborg}\ \emph {et~al.}(2014)\citenamefont {Jarlborg}, \citenamefont {Barbiellini}, \citenamefont {Lane}, \citenamefont {Wang}, \citenamefont {Markiewicz}, \citenamefont {Liu}, \citenamefont {Hussain},\ and\ \citenamefont {Bansil}}]{jarlborg2014}%
  \BibitemOpen
  \bibfield  {author} {\bibinfo {author} {\bibfnamefont {T.}~\bibnamefont {Jarlborg}}, \bibinfo {author} {\bibfnamefont {B.}~\bibnamefont {Barbiellini}}, \bibinfo {author} {\bibfnamefont {C.}~\bibnamefont {Lane}}, \bibinfo {author} {\bibfnamefont {Y.~J.}\ \bibnamefont {Wang}}, \bibinfo {author} {\bibfnamefont {R.}~\bibnamefont {Markiewicz}}, \bibinfo {author} {\bibfnamefont {Z.}~\bibnamefont {Liu}}, \bibinfo {author} {\bibfnamefont {Z.}~\bibnamefont {Hussain}}, \ and\ \bibinfo {author} {\bibfnamefont {A.}~\bibnamefont {Bansil}},\ }\bibfield  {title} {\enquote {\bibinfo {title} {{Electronic structure and excitations in oxygen deficient CeO$_{2-\delta}$ from DFT calculations}},}\ }\href@noop {} {\bibfield  {journal} {\bibinfo  {journal} {Physical Review B}\ }\textbf {\bibinfo {volume} {89}},\ \bibinfo {pages} {165101} (\bibinfo {year} {2014})}\BibitemShut {NoStop}%
\bibitem [{\citenamefont {Furness}\ \emph {et~al.}(2020{\natexlab{a}})\citenamefont {Furness}, \citenamefont {Kaplan}, \citenamefont {Ning}, \citenamefont {Perdew},\ and\ \citenamefont {Sun}}]{r2SCAN}%
  \BibitemOpen
  \bibfield  {author} {\bibinfo {author} {\bibfnamefont {J.~W.}\ \bibnamefont {Furness}}, \bibinfo {author} {\bibfnamefont {A.~D.}\ \bibnamefont {Kaplan}}, \bibinfo {author} {\bibfnamefont {J.}~\bibnamefont {Ning}}, \bibinfo {author} {\bibfnamefont {J.~P.}\ \bibnamefont {Perdew}}, \ and\ \bibinfo {author} {\bibfnamefont {J.}~\bibnamefont {Sun}},\ }\bibfield  {title} {\enquote {\bibinfo {title} {{Accurate and numerically efficient r2SCAN meta-generalized gradient approximation}},}\ }\href {\doibase 10.1021/acs.jpclett.0c02405} {\bibfield  {journal} {\bibinfo  {journal} {J. Phys. Chem. Lett.}\ }\textbf {\bibinfo {volume} {11}},\ \bibinfo {pages} {8208--8215} (\bibinfo {year} {2020}{\natexlab{a}})}\BibitemShut {NoStop}%
\bibitem [{\citenamefont {Ning}\ \emph {et~al.}(2023)\citenamefont {Ning}, \citenamefont {Lane}, \citenamefont {Zhang}, \citenamefont {Matzelle}, \citenamefont {Singh}, \citenamefont {Barbiellini}, \citenamefont {Markiewicz}, \citenamefont {Bansil},\ and\ \citenamefont {Sun}}]{YBCO6_NJL}%
  \BibitemOpen
  \bibfield  {author} {\bibinfo {author} {\bibfnamefont {J.}~\bibnamefont {Ning}}, \bibinfo {author} {\bibfnamefont {C.}~\bibnamefont {Lane}}, \bibinfo {author} {\bibfnamefont {Y.}~\bibnamefont {Zhang}}, \bibinfo {author} {\bibfnamefont {M.}~\bibnamefont {Matzelle}}, \bibinfo {author} {\bibfnamefont {B.}~\bibnamefont {Singh}}, \bibinfo {author} {\bibfnamefont {B.}~\bibnamefont {Barbiellini}}, \bibinfo {author} {\bibfnamefont {R.~S.}\ \bibnamefont {Markiewicz}}, \bibinfo {author} {\bibfnamefont {A.}~\bibnamefont {Bansil}}, \ and\ \bibinfo {author} {\bibfnamefont {J.}~\bibnamefont {Sun}},\ }\bibfield  {title} {\enquote {\bibinfo {title} {Critical role of magnetic moments in the lattice dynamics of ${\mathrm{yba}}_{2}{\mathrm{cu}}_{3}{\mathrm{o}}_{6}$},}\ }\href {\doibase 10.1103/PhysRevB.107.045126} {\bibfield  {journal} {\bibinfo  {journal} {Phys. Rev. B}\ }\textbf {\bibinfo {volume} {107}},\ \bibinfo {pages} {045126} (\bibinfo {year} {2023})}\BibitemShut {NoStop}%
\bibitem [{\citenamefont {Sun}, \citenamefont {Ruzsinszky},\ and\ \citenamefont {Perdew}(2015)}]{SCAN}%
  \BibitemOpen
  \bibfield  {author} {\bibinfo {author} {\bibfnamefont {J.}~\bibnamefont {Sun}}, \bibinfo {author} {\bibfnamefont {A.}~\bibnamefont {Ruzsinszky}}, \ and\ \bibinfo {author} {\bibfnamefont {J.~P.}\ \bibnamefont {Perdew}},\ }\bibfield  {title} {\enquote {\bibinfo {title} {Strongly constrained and appropriately normed semilocal density functional},}\ }\href {\doibase 10.1103/PhysRevLett.115.036402} {\bibfield  {journal} {\bibinfo  {journal} {Phys. Rev. Lett.}\ }\textbf {\bibinfo {volume} {115}},\ \bibinfo {pages} {036402} (\bibinfo {year} {2015})}\BibitemShut {NoStop}%
\bibitem [{\citenamefont {Sun}\ \emph {et~al.}(2016)\citenamefont {Sun}, \citenamefont {Remsing}, \citenamefont {Zhang}, \citenamefont {Sun}, \citenamefont {Ruzsinszky}, \citenamefont {Peng}, \citenamefont {Yang}, \citenamefont {Paul}, \citenamefont {Waghmare}, \citenamefont {Wu} \emph {et~al.}}]{SCAN_NChem}%
  \BibitemOpen
  \bibfield  {author} {\bibinfo {author} {\bibfnamefont {J.}~\bibnamefont {Sun}}, \bibinfo {author} {\bibfnamefont {R.~C.}\ \bibnamefont {Remsing}}, \bibinfo {author} {\bibfnamefont {Y.}~\bibnamefont {Zhang}}, \bibinfo {author} {\bibfnamefont {Z.}~\bibnamefont {Sun}}, \bibinfo {author} {\bibfnamefont {A.}~\bibnamefont {Ruzsinszky}}, \bibinfo {author} {\bibfnamefont {H.}~\bibnamefont {Peng}}, \bibinfo {author} {\bibfnamefont {Z.}~\bibnamefont {Yang}}, \bibinfo {author} {\bibfnamefont {A.}~\bibnamefont {Paul}}, \bibinfo {author} {\bibfnamefont {U.}~\bibnamefont {Waghmare}}, \bibinfo {author} {\bibfnamefont {X.}~\bibnamefont {Wu}},  \emph {et~al.},\ }\bibfield  {title} {\enquote {\bibinfo {title} {Accurate first-principles structures and energies of diversely bonded systems from an efficient density functional},}\ }\href@noop {} {\bibfield  {journal} {\bibinfo  {journal} {Nature chemistry}\ }\textbf {\bibinfo {volume} {8}},\ \bibinfo {pages} {831} (\bibinfo {year} {2016})}\BibitemShut {NoStop}%
\bibitem [{\citenamefont {Zhang}\ \emph {et~al.}(2020{\natexlab{b}})\citenamefont {Zhang}, \citenamefont {Furness}, \citenamefont {Zhang}, \citenamefont {Wang}, \citenamefont {Zunger},\ and\ \citenamefont {Sun}}]{yubo_MO}%
  \BibitemOpen
  \bibfield  {author} {\bibinfo {author} {\bibfnamefont {Y.}~\bibnamefont {Zhang}}, \bibinfo {author} {\bibfnamefont {J.}~\bibnamefont {Furness}}, \bibinfo {author} {\bibfnamefont {R.}~\bibnamefont {Zhang}}, \bibinfo {author} {\bibfnamefont {Z.}~\bibnamefont {Wang}}, \bibinfo {author} {\bibfnamefont {A.}~\bibnamefont {Zunger}}, \ and\ \bibinfo {author} {\bibfnamefont {J.}~\bibnamefont {Sun}},\ }\bibfield  {title} {\enquote {\bibinfo {title} {Symmetry-breaking polymorphous descriptions for correlated materials without interelectronic {U}},}\ }\href {\doibase 10.1103/PhysRevB.102.045112} {\bibfield  {journal} {\bibinfo  {journal} {Phys. Rev. B}\ }\textbf {\bibinfo {volume} {102}},\ \bibinfo {pages} {045112} (\bibinfo {year} {2020}{\natexlab{b}})}\BibitemShut {NoStop}%
\bibitem [{\citenamefont {Zhang}\ \emph {et~al.}(2019)\citenamefont {Zhang}, \citenamefont {Furness}, \citenamefont {Xiao},\ and\ \citenamefont {Sun}}]{Yubo_TiO2}%
  \BibitemOpen
  \bibfield  {author} {\bibinfo {author} {\bibfnamefont {Y.}~\bibnamefont {Zhang}}, \bibinfo {author} {\bibfnamefont {J.~W.}\ \bibnamefont {Furness}}, \bibinfo {author} {\bibfnamefont {B.}~\bibnamefont {Xiao}}, \ and\ \bibinfo {author} {\bibfnamefont {J.}~\bibnamefont {Sun}},\ }\bibfield  {title} {\enquote {\bibinfo {title} {Subtlety of {TiO$_2$} phase stability: Reliability of the density functional theory predictions and persistence of the self-interaction error},}\ }\href@noop {} {\bibfield  {journal} {\bibinfo  {journal} {The Journal of Chemical Physics}\ }\textbf {\bibinfo {volume} {150}},\ \bibinfo {pages} {014105} (\bibinfo {year} {2019})}\BibitemShut {NoStop}%
\bibitem [{\citenamefont {Kitchaev}\ \emph {et~al.}(2016)\citenamefont {Kitchaev}, \citenamefont {Peng}, \citenamefont {Liu}, \citenamefont {Sun}, \citenamefont {Perdew},\ and\ \citenamefont {Ceder}}]{Peng_MnO2}%
  \BibitemOpen
  \bibfield  {author} {\bibinfo {author} {\bibfnamefont {D.~A.}\ \bibnamefont {Kitchaev}}, \bibinfo {author} {\bibfnamefont {H.}~\bibnamefont {Peng}}, \bibinfo {author} {\bibfnamefont {Y.}~\bibnamefont {Liu}}, \bibinfo {author} {\bibfnamefont {J.}~\bibnamefont {Sun}}, \bibinfo {author} {\bibfnamefont {J.~P.}\ \bibnamefont {Perdew}}, \ and\ \bibinfo {author} {\bibfnamefont {G.}~\bibnamefont {Ceder}},\ }\bibfield  {title} {\enquote {\bibinfo {title} {Energetics of {${\mathrm{MnO}}_{2}$} polymorphs in density functional theory},}\ }\href {\doibase 10.1103/PhysRevB.93.045132} {\bibfield  {journal} {\bibinfo  {journal} {Phys. Rev. B}\ }\textbf {\bibinfo {volume} {93}},\ \bibinfo {pages} {045132} (\bibinfo {year} {2016})}\BibitemShut {NoStop}%
\bibitem [{\citenamefont {Peng}\ and\ \citenamefont {Perdew}(2017{\natexlab{a}})}]{Peng_MO}%
  \BibitemOpen
  \bibfield  {author} {\bibinfo {author} {\bibfnamefont {H.}~\bibnamefont {Peng}}\ and\ \bibinfo {author} {\bibfnamefont {J.~P.}\ \bibnamefont {Perdew}},\ }\bibfield  {title} {\enquote {\bibinfo {title} {Synergy of van der waals and self-interaction corrections in transition metal monoxides},}\ }\href {\doibase 10.1103/PhysRevB.96.100101} {\bibfield  {journal} {\bibinfo  {journal} {Phys. Rev. B}\ }\textbf {\bibinfo {volume} {96}},\ \bibinfo {pages} {100101} (\bibinfo {year} {2017}{\natexlab{a}})}\BibitemShut {NoStop}%
\bibitem [{\citenamefont {Ning}\ \emph {et~al.}(2020)\citenamefont {Ning}, \citenamefont {Zhu}, \citenamefont {Kidd}, \citenamefont {Guan}, \citenamefont {Wang}, \citenamefont {Mao},\ and\ \citenamefont {Sun}}]{ning_MBT}%
  \BibitemOpen
  \bibfield  {author} {\bibinfo {author} {\bibfnamefont {J.}~\bibnamefont {Ning}}, \bibinfo {author} {\bibfnamefont {Y.}~\bibnamefont {Zhu}}, \bibinfo {author} {\bibfnamefont {J.}~\bibnamefont {Kidd}}, \bibinfo {author} {\bibfnamefont {Y.}~\bibnamefont {Guan}}, \bibinfo {author} {\bibfnamefont {Y.}~\bibnamefont {Wang}}, \bibinfo {author} {\bibfnamefont {Z.}~\bibnamefont {Mao}}, \ and\ \bibinfo {author} {\bibfnamefont {J.}~\bibnamefont {Sun}},\ }\bibfield  {title} {\enquote {\bibinfo {title} {Subtle metastability of the layered magnetic topological insulator {MnBi$_{2}$Te$_{4}$} from weak interactions},}\ }\href@noop {} {\bibfield  {journal} {\bibinfo  {journal} {npj Computational Materials}\ }\textbf {\bibinfo {volume} {6}},\ \bibinfo {pages} {1--6} (\bibinfo {year} {2020})}\BibitemShut {NoStop}%
\bibitem [{\citenamefont {Lane}\ \emph {et~al.}(2020)\citenamefont {Lane}, \citenamefont {Zhang}, \citenamefont {Furness}, \citenamefont {Markiewicz}, \citenamefont {Barbiellini}, \citenamefont {Sun},\ and\ \citenamefont {Bansil}}]{SIO_Chris}%
  \BibitemOpen
  \bibfield  {author} {\bibinfo {author} {\bibfnamefont {C.}~\bibnamefont {Lane}}, \bibinfo {author} {\bibfnamefont {Y.}~\bibnamefont {Zhang}}, \bibinfo {author} {\bibfnamefont {J.~W.}\ \bibnamefont {Furness}}, \bibinfo {author} {\bibfnamefont {R.~S.}\ \bibnamefont {Markiewicz}}, \bibinfo {author} {\bibfnamefont {B.}~\bibnamefont {Barbiellini}}, \bibinfo {author} {\bibfnamefont {J.}~\bibnamefont {Sun}}, \ and\ \bibinfo {author} {\bibfnamefont {A.}~\bibnamefont {Bansil}},\ }\bibfield  {title} {\enquote {\bibinfo {title} {First-principles calculation of spin and orbital contributions to magnetically ordered moments in {${\mathrm{Sr}}_{2}{\mathrm{IrO}}_{4}$}},}\ }\href {\doibase 10.1103/PhysRevB.101.155110} {\bibfield  {journal} {\bibinfo  {journal} {Phys. Rev. B}\ }\textbf {\bibinfo {volume} {101}},\ \bibinfo {pages} {155110} (\bibinfo {year} {2020})}\BibitemShut {NoStop}%
\bibitem [{\citenamefont {Pokharel}\ \emph {et~al.}(2022)\citenamefont {Pokharel}, \citenamefont {Lane}, \citenamefont {Furness}, \citenamefont {Zhang}, \citenamefont {Ning}, \citenamefont {Barbiellini}, \citenamefont {Markiewicz}, \citenamefont {Zhang}, \citenamefont {Bansil},\ and\ \citenamefont {Sun}}]{Kanun_NPJ}%
  \BibitemOpen
  \bibfield  {author} {\bibinfo {author} {\bibfnamefont {K.}~\bibnamefont {Pokharel}}, \bibinfo {author} {\bibfnamefont {C.}~\bibnamefont {Lane}}, \bibinfo {author} {\bibfnamefont {J.~W.}\ \bibnamefont {Furness}}, \bibinfo {author} {\bibfnamefont {R.}~\bibnamefont {Zhang}}, \bibinfo {author} {\bibfnamefont {J.}~\bibnamefont {Ning}}, \bibinfo {author} {\bibfnamefont {B.}~\bibnamefont {Barbiellini}}, \bibinfo {author} {\bibfnamefont {R.~S.}\ \bibnamefont {Markiewicz}}, \bibinfo {author} {\bibfnamefont {Y.}~\bibnamefont {Zhang}}, \bibinfo {author} {\bibfnamefont {A.}~\bibnamefont {Bansil}}, \ and\ \bibinfo {author} {\bibfnamefont {J.}~\bibnamefont {Sun}},\ }\bibfield  {title} {\enquote {\bibinfo {title} {Sensitivity of the electronic and magnetic structures of cuprate superconductors to density functional approximations},}\ }\href@noop {} {\bibfield  {journal} {\bibinfo  {journal} {npj Computational Materials}\ }\textbf {\bibinfo {volume} {8}},\ \bibinfo {pages} {1--11} (\bibinfo {year} {2022})}\BibitemShut
  {NoStop}%
\bibitem [{\citenamefont {Zhang}\ \emph {et~al.}(2022)\citenamefont {Zhang}, \citenamefont {Singh}, \citenamefont {Lane}, \citenamefont {Kidd}, \citenamefont {Zhang}, \citenamefont {Barbiellini}, \citenamefont {Markiewicz}, \citenamefont {Bansil},\ and\ \citenamefont {Sun}}]{SmB6}%
  \BibitemOpen
  \bibfield  {author} {\bibinfo {author} {\bibfnamefont {R.}~\bibnamefont {Zhang}}, \bibinfo {author} {\bibfnamefont {B.}~\bibnamefont {Singh}}, \bibinfo {author} {\bibfnamefont {C.}~\bibnamefont {Lane}}, \bibinfo {author} {\bibfnamefont {J.}~\bibnamefont {Kidd}}, \bibinfo {author} {\bibfnamefont {Y.}~\bibnamefont {Zhang}}, \bibinfo {author} {\bibfnamefont {B.}~\bibnamefont {Barbiellini}}, \bibinfo {author} {\bibfnamefont {R.~S.}\ \bibnamefont {Markiewicz}}, \bibinfo {author} {\bibfnamefont {A.}~\bibnamefont {Bansil}}, \ and\ \bibinfo {author} {\bibfnamefont {J.}~\bibnamefont {Sun}},\ }\bibfield  {title} {\enquote {\bibinfo {title} {Critical role of magnetic moments in heavy-fermion materials: Revisiting ${\mathrm{smb}}_{6}$},}\ }\href {\doibase 10.1103/PhysRevB.105.195134} {\bibfield  {journal} {\bibinfo  {journal} {Phys. Rev. B}\ }\textbf {\bibinfo {volume} {105}},\ \bibinfo {pages} {195134} (\bibinfo {year} {2022})}\BibitemShut {NoStop}%
\bibitem [{\citenamefont {Ning}, \citenamefont {Furness},\ and\ \citenamefont {Sun}(2022)}]{r2SCAN_phonon}%
  \BibitemOpen
  \bibfield  {author} {\bibinfo {author} {\bibfnamefont {J.}~\bibnamefont {Ning}}, \bibinfo {author} {\bibfnamefont {J.~W.}\ \bibnamefont {Furness}}, \ and\ \bibinfo {author} {\bibfnamefont {J.}~\bibnamefont {Sun}},\ }\bibfield  {title} {\enquote {\bibinfo {title} {Reliable lattice dynamics from an efficient density functional approximation},}\ }\href@noop {} {\bibfield  {journal} {\bibinfo  {journal} {Chemistry of Materials}\ }\textbf {\bibinfo {volume} {34}},\ \bibinfo {pages} {2562--2568} (\bibinfo {year} {2022})}\BibitemShut {NoStop}%
\bibitem [{\citenamefont {Ning}\ \emph {et~al.}(2022)\citenamefont {Ning}, \citenamefont {Kothakonda}, \citenamefont {Furness}, \citenamefont {Kaplan}, \citenamefont {Ehlert}, \citenamefont {Brandenburg}, \citenamefont {Perdew},\ and\ \citenamefont {Sun}}]{r2SCAN_rVV10}%
  \BibitemOpen
  \bibfield  {author} {\bibinfo {author} {\bibfnamefont {J.}~\bibnamefont {Ning}}, \bibinfo {author} {\bibfnamefont {M.}~\bibnamefont {Kothakonda}}, \bibinfo {author} {\bibfnamefont {J.~W.}\ \bibnamefont {Furness}}, \bibinfo {author} {\bibfnamefont {A.~D.}\ \bibnamefont {Kaplan}}, \bibinfo {author} {\bibfnamefont {S.}~\bibnamefont {Ehlert}}, \bibinfo {author} {\bibfnamefont {J.~G.}\ \bibnamefont {Brandenburg}}, \bibinfo {author} {\bibfnamefont {J.~P.}\ \bibnamefont {Perdew}}, \ and\ \bibinfo {author} {\bibfnamefont {J.}~\bibnamefont {Sun}},\ }\bibfield  {title} {\enquote {\bibinfo {title} {Workhorse minimally empirical dispersion-corrected density functional with tests for weakly bound systems: r2scan+rvv10},}\ }\href {\doibase 10.1103/PhysRevB.106.075422} {\bibfield  {journal} {\bibinfo  {journal} {Phys. Rev. B}\ }\textbf {\bibinfo {volume} {106}},\ \bibinfo {pages} {075422} (\bibinfo {year} {2022})}\BibitemShut {NoStop}%
\bibitem [{\citenamefont {Mathis}\ \emph {et~al.}(2022)\citenamefont {Mathis}, \citenamefont {Khanolkar}, \citenamefont {Fu}, \citenamefont {Bryan}, \citenamefont {Dennett}, \citenamefont {Rickert}, \citenamefont {Mann}, \citenamefont {Winn}, \citenamefont {Abernathy}, \citenamefont {Manley}, \citenamefont {Hurley},\ and\ \citenamefont {Marianetti}}]{GQHA1}%
  \BibitemOpen
  \bibfield  {author} {\bibinfo {author} {\bibfnamefont {M.~A.}\ \bibnamefont {Mathis}}, \bibinfo {author} {\bibfnamefont {A.}~\bibnamefont {Khanolkar}}, \bibinfo {author} {\bibfnamefont {L.}~\bibnamefont {Fu}}, \bibinfo {author} {\bibfnamefont {M.~S.}\ \bibnamefont {Bryan}}, \bibinfo {author} {\bibfnamefont {C.~A.}\ \bibnamefont {Dennett}}, \bibinfo {author} {\bibfnamefont {K.}~\bibnamefont {Rickert}}, \bibinfo {author} {\bibfnamefont {J.~M.}\ \bibnamefont {Mann}}, \bibinfo {author} {\bibfnamefont {B.}~\bibnamefont {Winn}}, \bibinfo {author} {\bibfnamefont {D.~L.}\ \bibnamefont {Abernathy}}, \bibinfo {author} {\bibfnamefont {M.~E.}\ \bibnamefont {Manley}}, \bibinfo {author} {\bibfnamefont {D.~H.}\ \bibnamefont {Hurley}}, \ and\ \bibinfo {author} {\bibfnamefont {C.~A.}\ \bibnamefont {Marianetti}},\ }\bibfield  {title} {\enquote {\bibinfo {title} {Generalized quasiharmonic approximation via space group irreducible derivatives},}\ }\href {\doibase 10.1103/PhysRevB.106.014314} {\bibfield  {journal} {\bibinfo
  {journal} {Phys. Rev. B}\ }\textbf {\bibinfo {volume} {106}},\ \bibinfo {pages} {014314} (\bibinfo {year} {2022})}\BibitemShut {NoStop}%
\bibitem [{\citenamefont {Xiao}\ \emph {et~al.}(2022)\citenamefont {Xiao}, \citenamefont {Ma}, \citenamefont {Bryan}, \citenamefont {Fu}, \citenamefont {Mann}, \citenamefont {Winn}, \citenamefont {Abernathy}, \citenamefont {Hermann}, \citenamefont {Khanolkar}, \citenamefont {Dennett}, \citenamefont {Hurley}, \citenamefont {Manley},\ and\ \citenamefont {Marianetti}}]{GQHA2}%
  \BibitemOpen
  \bibfield  {author} {\bibinfo {author} {\bibfnamefont {E.}~\bibnamefont {Xiao}}, \bibinfo {author} {\bibfnamefont {H.}~\bibnamefont {Ma}}, \bibinfo {author} {\bibfnamefont {M.~S.}\ \bibnamefont {Bryan}}, \bibinfo {author} {\bibfnamefont {L.}~\bibnamefont {Fu}}, \bibinfo {author} {\bibfnamefont {J.~M.}\ \bibnamefont {Mann}}, \bibinfo {author} {\bibfnamefont {B.}~\bibnamefont {Winn}}, \bibinfo {author} {\bibfnamefont {D.~L.}\ \bibnamefont {Abernathy}}, \bibinfo {author} {\bibfnamefont {R.~P.}\ \bibnamefont {Hermann}}, \bibinfo {author} {\bibfnamefont {A.~R.}\ \bibnamefont {Khanolkar}}, \bibinfo {author} {\bibfnamefont {C.~A.}\ \bibnamefont {Dennett}}, \bibinfo {author} {\bibfnamefont {D.~H.}\ \bibnamefont {Hurley}}, \bibinfo {author} {\bibfnamefont {M.~E.}\ \bibnamefont {Manley}}, \ and\ \bibinfo {author} {\bibfnamefont {C.~A.}\ \bibnamefont {Marianetti}},\ }\bibfield  {title} {\enquote {\bibinfo {title} {Validating first-principles phonon lifetimes via inelastic neutron scattering},}\ }\href {\doibase
  10.1103/PhysRevB.106.144310} {\bibfield  {journal} {\bibinfo  {journal} {Phys. Rev. B}\ }\textbf {\bibinfo {volume} {106}},\ \bibinfo {pages} {144310} (\bibinfo {year} {2022})}\BibitemShut {NoStop}%
\bibitem [{\citenamefont {Bandi}\ and\ \citenamefont {Marianetti}(2023)}]{phonon_irr_der}%
  \BibitemOpen
  \bibfield  {author} {\bibinfo {author} {\bibfnamefont {S.}~\bibnamefont {Bandi}}\ and\ \bibinfo {author} {\bibfnamefont {C.~A.}\ \bibnamefont {Marianetti}},\ }\bibfield  {title} {\enquote {\bibinfo {title} {Precisely computing phonons via irreducible derivatives},}\ }\href {\doibase 10.1103/PhysRevB.107.174302} {\bibfield  {journal} {\bibinfo  {journal} {Phys. Rev. B}\ }\textbf {\bibinfo {volume} {107}},\ \bibinfo {pages} {174302} (\bibinfo {year} {2023})}\BibitemShut {NoStop}%
\bibitem [{\citenamefont {Perdew}\ and\ \citenamefont {Zunger}(1981)}]{PZSIC}%
  \BibitemOpen
  \bibfield  {author} {\bibinfo {author} {\bibfnamefont {J.~P.}\ \bibnamefont {Perdew}}\ and\ \bibinfo {author} {\bibfnamefont {A.}~\bibnamefont {Zunger}},\ }\bibfield  {title} {\enquote {\bibinfo {title} {Self-interaction correction to density-functional approximations for many-electron systems},}\ }\href {\doibase 10.1103/PhysRevB.23.5048} {\bibfield  {journal} {\bibinfo  {journal} {Phys. Rev. B}\ }\textbf {\bibinfo {volume} {23}},\ \bibinfo {pages} {5048--5079} (\bibinfo {year} {1981})}\BibitemShut {NoStop}%
\bibitem [{\citenamefont {Falter}, \citenamefont {Klenner},\ and\ \citenamefont {Ludwig}(1993)}]{model_Falter1}%
  \BibitemOpen
  \bibfield  {author} {\bibinfo {author} {\bibfnamefont {C.}~\bibnamefont {Falter}}, \bibinfo {author} {\bibfnamefont {M.}~\bibnamefont {Klenner}}, \ and\ \bibinfo {author} {\bibfnamefont {W.}~\bibnamefont {Ludwig}},\ }\bibfield  {title} {\enquote {\bibinfo {title} {Effect of charge fluctuations on the phonon dispersion and electron-phonon interaction in {${\mathrm{La}}_{2}$${\mathrm{CuO}}_{4}$}},}\ }\href {\doibase 10.1103/PhysRevB.47.5390} {\bibfield  {journal} {\bibinfo  {journal} {Phys. Rev. B}\ }\textbf {\bibinfo {volume} {47}},\ \bibinfo {pages} {5390--5404} (\bibinfo {year} {1993})}\BibitemShut {NoStop}%
\bibitem [{\citenamefont {Zhou}\ \emph {et~al.}(2022)\citenamefont {Zhou}, \citenamefont {Ma}, \citenamefont {Xiao}, \citenamefont {Gofryk}, \citenamefont {Jiang}, \citenamefont {Manley}, \citenamefont {Hurley},\ and\ \citenamefont {Marianetti}}]{UO2_phonon}%
  \BibitemOpen
  \bibfield  {author} {\bibinfo {author} {\bibfnamefont {S.}~\bibnamefont {Zhou}}, \bibinfo {author} {\bibfnamefont {H.}~\bibnamefont {Ma}}, \bibinfo {author} {\bibfnamefont {E.}~\bibnamefont {Xiao}}, \bibinfo {author} {\bibfnamefont {K.}~\bibnamefont {Gofryk}}, \bibinfo {author} {\bibfnamefont {C.}~\bibnamefont {Jiang}}, \bibinfo {author} {\bibfnamefont {M.~E.}\ \bibnamefont {Manley}}, \bibinfo {author} {\bibfnamefont {D.~H.}\ \bibnamefont {Hurley}}, \ and\ \bibinfo {author} {\bibfnamefont {C.~A.}\ \bibnamefont {Marianetti}},\ }\bibfield  {title} {\enquote {\bibinfo {title} {Capturing the ground state of uranium dioxide from first principles: Crystal distortion, magnetic structure, and phonons},}\ }\href {\doibase 10.1103/PhysRevB.106.125134} {\bibfield  {journal} {\bibinfo  {journal} {Phys. Rev. B}\ }\textbf {\bibinfo {volume} {106}},\ \bibinfo {pages} {125134} (\bibinfo {year} {2022})}\BibitemShut {NoStop}%
\bibitem [{\citenamefont {Furness}\ \emph {et~al.}(2022)\citenamefont {Furness}, \citenamefont {Kaplan}, \citenamefont {Ning}, \citenamefont {Perdew},\ and\ \citenamefont {Sun}}]{r4SCAN}%
  \BibitemOpen
  \bibfield  {author} {\bibinfo {author} {\bibfnamefont {J.~W.}\ \bibnamefont {Furness}}, \bibinfo {author} {\bibfnamefont {A.~D.}\ \bibnamefont {Kaplan}}, \bibinfo {author} {\bibfnamefont {J.}~\bibnamefont {Ning}}, \bibinfo {author} {\bibfnamefont {J.~P.}\ \bibnamefont {Perdew}}, \ and\ \bibinfo {author} {\bibfnamefont {J.}~\bibnamefont {Sun}},\ }\bibfield  {title} {\enquote {\bibinfo {title} {{Construction of meta-GGA functionals through restoration of exact constraint adherence to regularized SCAN functionals}},}\ }\href@noop {} {\bibfield  {journal} {\bibinfo  {journal} {The Journal of Chemical Physics}\ }\textbf {\bibinfo {volume} {156}},\ \bibinfo {pages} {034109} (\bibinfo {year} {2022})}\BibitemShut {NoStop}%
\bibitem [{\citenamefont {Kaplan}, \citenamefont {Levy},\ and\ \citenamefont {Perdew}(2023)}]{Aaron_review}%
  \BibitemOpen
  \bibfield  {author} {\bibinfo {author} {\bibfnamefont {A.~D.}\ \bibnamefont {Kaplan}}, \bibinfo {author} {\bibfnamefont {M.}~\bibnamefont {Levy}}, \ and\ \bibinfo {author} {\bibfnamefont {J.~P.}\ \bibnamefont {Perdew}},\ }\bibfield  {title} {\enquote {\bibinfo {title} {The predictive power of exact constraints and appropriate norms in density functional theory},}\ }\href@noop {} {\bibfield  {journal} {\bibinfo  {journal} {Annual Review of Physical Chemistry}\ }\textbf {\bibinfo {volume} {74}},\ \bibinfo {pages} {193--218} (\bibinfo {year} {2023})}\BibitemShut {NoStop}%
\bibitem [{\citenamefont {Tao}\ \emph {et~al.}(2017)\citenamefont {Tao}, \citenamefont {Zheng}, \citenamefont {Gebhardt}, \citenamefont {Perdew},\ and\ \citenamefont {Rappe}}]{vdw_solids_Tao}%
  \BibitemOpen
  \bibfield  {author} {\bibinfo {author} {\bibfnamefont {J.}~\bibnamefont {Tao}}, \bibinfo {author} {\bibfnamefont {F.}~\bibnamefont {Zheng}}, \bibinfo {author} {\bibfnamefont {J.}~\bibnamefont {Gebhardt}}, \bibinfo {author} {\bibfnamefont {J.~P.}\ \bibnamefont {Perdew}}, \ and\ \bibinfo {author} {\bibfnamefont {A.~M.}\ \bibnamefont {Rappe}},\ }\bibfield  {title} {\enquote {\bibinfo {title} {Screened van der waals correction to density functional theory for solids},}\ }\href {\doibase 10.1103/PhysRevMaterials.1.020802} {\bibfield  {journal} {\bibinfo  {journal} {Phys. Rev. Mater.}\ }\textbf {\bibinfo {volume} {1}},\ \bibinfo {pages} {020802} (\bibinfo {year} {2017})}\BibitemShut {NoStop}%
\bibitem [{\citenamefont {Caldeweyher}, \citenamefont {Bannwarth},\ and\ \citenamefont {Grimme}(2017)}]{D4}%
  \BibitemOpen
  \bibfield  {author} {\bibinfo {author} {\bibfnamefont {E.}~\bibnamefont {Caldeweyher}}, \bibinfo {author} {\bibfnamefont {C.}~\bibnamefont {Bannwarth}}, \ and\ \bibinfo {author} {\bibfnamefont {S.}~\bibnamefont {Grimme}},\ }\bibfield  {title} {\enquote {\bibinfo {title} {{Extension of the D3 dispersion coefficient model}},}\ }\href@noop {} {\bibfield  {journal} {\bibinfo  {journal} {The Journal of Chemical Physics}\ }\textbf {\bibinfo {volume} {147}},\ \bibinfo {pages} {034112} (\bibinfo {year} {2017})}\BibitemShut {NoStop}%
\bibitem [{\citenamefont {Caldeweyher}\ \emph {et~al.}(2020)\citenamefont {Caldeweyher}, \citenamefont {Mewes}, \citenamefont {Ehlert},\ and\ \citenamefont {Grimme}}]{D4_test}%
  \BibitemOpen
  \bibfield  {author} {\bibinfo {author} {\bibfnamefont {E.}~\bibnamefont {Caldeweyher}}, \bibinfo {author} {\bibfnamefont {J.-M.}\ \bibnamefont {Mewes}}, \bibinfo {author} {\bibfnamefont {S.}~\bibnamefont {Ehlert}}, \ and\ \bibinfo {author} {\bibfnamefont {S.}~\bibnamefont {Grimme}},\ }\bibfield  {title} {\enquote {\bibinfo {title} {Extension and evaluation of the d4 london-dispersion model for periodic systems},}\ }\href@noop {} {\bibfield  {journal} {\bibinfo  {journal} {Physical Chemistry Chemical Physics}\ }\textbf {\bibinfo {volume} {22}},\ \bibinfo {pages} {8499--8512} (\bibinfo {year} {2020})}\BibitemShut {NoStop}%
\bibitem [{\citenamefont {Caldeweyher}\ \emph {et~al.}(2019)\citenamefont {Caldeweyher}, \citenamefont {Ehlert}, \citenamefont {Hansen}, \citenamefont {Neugebauer}, \citenamefont {Spicher}, \citenamefont {Bannwarth},\ and\ \citenamefont {Grimme}}]{D4_parameter}%
  \BibitemOpen
  \bibfield  {author} {\bibinfo {author} {\bibfnamefont {E.}~\bibnamefont {Caldeweyher}}, \bibinfo {author} {\bibfnamefont {S.}~\bibnamefont {Ehlert}}, \bibinfo {author} {\bibfnamefont {A.}~\bibnamefont {Hansen}}, \bibinfo {author} {\bibfnamefont {H.}~\bibnamefont {Neugebauer}}, \bibinfo {author} {\bibfnamefont {S.}~\bibnamefont {Spicher}}, \bibinfo {author} {\bibfnamefont {C.}~\bibnamefont {Bannwarth}}, \ and\ \bibinfo {author} {\bibfnamefont {S.}~\bibnamefont {Grimme}},\ }\bibfield  {title} {\enquote {\bibinfo {title} {{A generally applicable atomic-charge dependent London dispersion correction}},}\ }\href@noop {} {\bibfield  {journal} {\bibinfo  {journal} {The Journal of Chemical Physics}\ }\textbf {\bibinfo {volume} {150}},\ \bibinfo {pages} {154122} (\bibinfo {year} {2019})}\BibitemShut {NoStop}%
\bibitem [{\citenamefont {Ehlert}\ \emph {et~al.}(2021)\citenamefont {Ehlert}, \citenamefont {Huniar}, \citenamefont {Ning}, \citenamefont {Furness}, \citenamefont {Sun}, \citenamefont {Kaplan}, \citenamefont {Perdew},\ and\ \citenamefont {Brandenburg}}]{r2SCAN_D4}%
  \BibitemOpen
  \bibfield  {author} {\bibinfo {author} {\bibfnamefont {S.}~\bibnamefont {Ehlert}}, \bibinfo {author} {\bibfnamefont {U.}~\bibnamefont {Huniar}}, \bibinfo {author} {\bibfnamefont {J.}~\bibnamefont {Ning}}, \bibinfo {author} {\bibfnamefont {J.~W.}\ \bibnamefont {Furness}}, \bibinfo {author} {\bibfnamefont {J.}~\bibnamefont {Sun}}, \bibinfo {author} {\bibfnamefont {A.~D.}\ \bibnamefont {Kaplan}}, \bibinfo {author} {\bibfnamefont {J.~P.}\ \bibnamefont {Perdew}}, \ and\ \bibinfo {author} {\bibfnamefont {J.~G.}\ \bibnamefont {Brandenburg}},\ }\bibfield  {title} {\enquote {\bibinfo {title} {{r2SCAN-D4: Dispersion corrected meta-generalized gradient approximation for general chemical applications}},}\ }\href@noop {} {\bibfield  {journal} {\bibinfo  {journal} {The Journal of Chemical Physics}\ }\textbf {\bibinfo {volume} {154}},\ \bibinfo {pages} {061101} (\bibinfo {year} {2021})}\BibitemShut {NoStop}%
\bibitem [{\citenamefont {Tao}\ and\ \citenamefont {Mo}(2016)}]{TM}%
  \BibitemOpen
  \bibfield  {author} {\bibinfo {author} {\bibfnamefont {J.}~\bibnamefont {Tao}}\ and\ \bibinfo {author} {\bibfnamefont {Y.}~\bibnamefont {Mo}},\ }\bibfield  {title} {\enquote {\bibinfo {title} {Accurate semilocal density functional for condensed-matter physics and quantum chemistry},}\ }\href {\doibase 10.1103/PhysRevLett.117.073001} {\bibfield  {journal} {\bibinfo  {journal} {Phys. Rev. Lett.}\ }\textbf {\bibinfo {volume} {117}},\ \bibinfo {pages} {073001} (\bibinfo {year} {2016})}\BibitemShut {NoStop}%
\bibitem [{\citenamefont {Jana}, \citenamefont {Sharma},\ and\ \citenamefont {Samal}(2019)}]{revTM}%
  \BibitemOpen
  \bibfield  {author} {\bibinfo {author} {\bibfnamefont {S.}~\bibnamefont {Jana}}, \bibinfo {author} {\bibfnamefont {K.}~\bibnamefont {Sharma}}, \ and\ \bibinfo {author} {\bibfnamefont {P.}~\bibnamefont {Samal}},\ }\bibfield  {title} {\enquote {\bibinfo {title} {Improving the performance of tao-mo non-empirical density functional with broader applicability in quantum chemistry and materials science},}\ }\href@noop {} {\bibfield  {journal} {\bibinfo  {journal} {The Journal of Physical Chemistry A}\ }\textbf {\bibinfo {volume} {123}},\ \bibinfo {pages} {6356--6369} (\bibinfo {year} {2019})}\BibitemShut {NoStop}%
\bibitem [{\citenamefont {Jana}\ \emph {et~al.}(2021)\citenamefont {Jana}, \citenamefont {Behera}, \citenamefont {{\'S}miga}, \citenamefont {Constantin},\ and\ \citenamefont {Samal}}]{rregTM}%
  \BibitemOpen
  \bibfield  {author} {\bibinfo {author} {\bibfnamefont {S.}~\bibnamefont {Jana}}, \bibinfo {author} {\bibfnamefont {S.~K.}\ \bibnamefont {Behera}}, \bibinfo {author} {\bibfnamefont {S.}~\bibnamefont {{\'S}miga}}, \bibinfo {author} {\bibfnamefont {L.~A.}\ \bibnamefont {Constantin}}, \ and\ \bibinfo {author} {\bibfnamefont {P.}~\bibnamefont {Samal}},\ }\bibfield  {title} {\enquote {\bibinfo {title} {Accurate density functional made more versatile},}\ }\href@noop {} {\bibfield  {journal} {\bibinfo  {journal} {The Journal of Chemical Physics}\ }\textbf {\bibinfo {volume} {155}} (\bibinfo {year} {2021})}\BibitemShut {NoStop}%
\bibitem [{\citenamefont {Furness}\ \emph {et~al.}(2020{\natexlab{b}})\citenamefont {Furness}, \citenamefont {Sengupta}, \citenamefont {Ning}, \citenamefont {Ruzsinszky},\ and\ \citenamefont {Sun}}]{OOL_TM}%
  \BibitemOpen
  \bibfield  {author} {\bibinfo {author} {\bibfnamefont {J.~W.}\ \bibnamefont {Furness}}, \bibinfo {author} {\bibfnamefont {N.}~\bibnamefont {Sengupta}}, \bibinfo {author} {\bibfnamefont {J.}~\bibnamefont {Ning}}, \bibinfo {author} {\bibfnamefont {A.}~\bibnamefont {Ruzsinszky}}, \ and\ \bibinfo {author} {\bibfnamefont {J.}~\bibnamefont {Sun}},\ }\bibfield  {title} {\enquote {\bibinfo {title} {{Examining the order-of-limits problem and lattice constant performance of the Tao-Mo functional}},}\ }\href {\doibase 10.1063/5.0008014} {\bibfield  {journal} {\bibinfo  {journal} {The Journal of Chemical Physics}\ }\textbf {\bibinfo {volume} {152}},\ \bibinfo {pages} {244112} (\bibinfo {year} {2020}{\natexlab{b}})}\BibitemShut {NoStop}%
\bibitem [{\citenamefont {Bl\"ochl}(1994)}]{PAW}%
  \BibitemOpen
  \bibfield  {author} {\bibinfo {author} {\bibfnamefont {P.~E.}\ \bibnamefont {Bl\"ochl}},\ }\bibfield  {title} {\enquote {\bibinfo {title} {Projector augmented-wave method},}\ }\href {\doibase 10.1103/PhysRevB.50.17953} {\bibfield  {journal} {\bibinfo  {journal} {Phys. Rev. B}\ }\textbf {\bibinfo {volume} {50}},\ \bibinfo {pages} {17953--17979} (\bibinfo {year} {1994})}\BibitemShut {NoStop}%
\bibitem [{\citenamefont {Kresse}\ and\ \citenamefont {Joubert}(1999)}]{PAW_vasp}%
  \BibitemOpen
  \bibfield  {author} {\bibinfo {author} {\bibfnamefont {G.}~\bibnamefont {Kresse}}\ and\ \bibinfo {author} {\bibfnamefont {D.}~\bibnamefont {Joubert}},\ }\bibfield  {title} {\enquote {\bibinfo {title} {From ultrasoft pseudopotentials to the projector augmented-wave method},}\ }\href@noop {} {\bibfield  {journal} {\bibinfo  {journal} {Phys. Rev. B}\ }\textbf {\bibinfo {volume} {59}},\ \bibinfo {pages} {1758} (\bibinfo {year} {1999})}\BibitemShut {NoStop}%
\bibitem [{\citenamefont {Kresse}\ and\ \citenamefont {Furthm\"uller}(1996)}]{VASP}%
  \BibitemOpen
  \bibfield  {author} {\bibinfo {author} {\bibfnamefont {G.}~\bibnamefont {Kresse}}\ and\ \bibinfo {author} {\bibfnamefont {J.}~\bibnamefont {Furthm\"uller}},\ }\bibfield  {title} {\enquote {\bibinfo {title} {Efficient iterative schemes for ab initio total-energy calculations using a plane-wave basis set},}\ }\href {\doibase 10.1103/PhysRevB.54.11169} {\bibfield  {journal} {\bibinfo  {journal} {Phys. Rev. B}\ }\textbf {\bibinfo {volume} {54}},\ \bibinfo {pages} {11169--11186} (\bibinfo {year} {1996})}\BibitemShut {NoStop}%
\bibitem [{\citenamefont {Kresse}\ and\ \citenamefont {Hafner}(1993)}]{VASP2}%
  \BibitemOpen
  \bibfield  {author} {\bibinfo {author} {\bibfnamefont {G.}~\bibnamefont {Kresse}}\ and\ \bibinfo {author} {\bibfnamefont {J.}~\bibnamefont {Hafner}},\ }\bibfield  {title} {\enquote {\bibinfo {title} {Ab initio molecular dynamics for open-shell transition metals},}\ }\href@noop {} {\bibfield  {journal} {\bibinfo  {journal} {Physical Review B}\ }\textbf {\bibinfo {volume} {48}},\ \bibinfo {pages} {13115} (\bibinfo {year} {1993})}\BibitemShut {NoStop}%
\bibitem [{\citenamefont {Togo}\ and\ \citenamefont {Tanaka}(2015)}]{phonopy}%
  \BibitemOpen
  \bibfield  {author} {\bibinfo {author} {\bibfnamefont {A.}~\bibnamefont {Togo}}\ and\ \bibinfo {author} {\bibfnamefont {I.}~\bibnamefont {Tanaka}},\ }\bibfield  {title} {\enquote {\bibinfo {title} {First principles phonon calculations in materials science},}\ }\href@noop {} {\bibfield  {journal} {\bibinfo  {journal} {Scripta Materialia}\ }\textbf {\bibinfo {volume} {108}},\ \bibinfo {pages} {1--5} (\bibinfo {year} {2015})}\BibitemShut {NoStop}%
\bibitem [{\citenamefont {Cava}\ \emph {et~al.}(1990)\citenamefont {Cava}, \citenamefont {Hewat}, \citenamefont {Hewat}, \citenamefont {Batlogg}, \citenamefont {Marezio}, \citenamefont {Rabe}, \citenamefont {Krajewski}, \citenamefont {Peck~Jr},\ and\ \citenamefont {Rupp~Jr}}]{ybco6_LC_expt}%
  \BibitemOpen
  \bibfield  {author} {\bibinfo {author} {\bibfnamefont {R.}~\bibnamefont {Cava}}, \bibinfo {author} {\bibfnamefont {A.}~\bibnamefont {Hewat}}, \bibinfo {author} {\bibfnamefont {E.}~\bibnamefont {Hewat}}, \bibinfo {author} {\bibfnamefont {B.}~\bibnamefont {Batlogg}}, \bibinfo {author} {\bibfnamefont {M.}~\bibnamefont {Marezio}}, \bibinfo {author} {\bibfnamefont {K.}~\bibnamefont {Rabe}}, \bibinfo {author} {\bibfnamefont {J.}~\bibnamefont {Krajewski}}, \bibinfo {author} {\bibfnamefont {W.}~\bibnamefont {Peck~Jr}}, \ and\ \bibinfo {author} {\bibfnamefont {L.}~\bibnamefont {Rupp~Jr}},\ }\bibfield  {title} {\enquote {\bibinfo {title} {Structural anomalies, oxygen ordering and superconductivity in oxygen deficient {Ba$_{2}$YCu$_{3}$O$_{x}$}},}\ }\href@noop {} {\bibfield  {journal} {\bibinfo  {journal} {Physica C: Superconductivity}\ }\textbf {\bibinfo {volume} {165}},\ \bibinfo {pages} {419--433} (\bibinfo {year} {1990})}\BibitemShut {NoStop}%
\bibitem [{\citenamefont {Casalta}\ \emph {et~al.}(1994)\citenamefont {Casalta}, \citenamefont {Schleger}, \citenamefont {Brecht}, \citenamefont {Montfrooij}, \citenamefont {Andersen}, \citenamefont {Lebech}, \citenamefont {Schmahl}, \citenamefont {Fuess}, \citenamefont {Liang}, \citenamefont {Hardy},\ and\ \citenamefont {Wolf}}]{casalta1994absence}%
  \BibitemOpen
  \bibfield  {author} {\bibinfo {author} {\bibfnamefont {H.}~\bibnamefont {Casalta}}, \bibinfo {author} {\bibfnamefont {P.}~\bibnamefont {Schleger}}, \bibinfo {author} {\bibfnamefont {E.}~\bibnamefont {Brecht}}, \bibinfo {author} {\bibfnamefont {W.}~\bibnamefont {Montfrooij}}, \bibinfo {author} {\bibfnamefont {N.~H.}\ \bibnamefont {Andersen}}, \bibinfo {author} {\bibfnamefont {B.}~\bibnamefont {Lebech}}, \bibinfo {author} {\bibfnamefont {W.~W.}\ \bibnamefont {Schmahl}}, \bibinfo {author} {\bibfnamefont {H.}~\bibnamefont {Fuess}}, \bibinfo {author} {\bibfnamefont {R.}~\bibnamefont {Liang}}, \bibinfo {author} {\bibfnamefont {W.~N.}\ \bibnamefont {Hardy}}, \ and\ \bibinfo {author} {\bibfnamefont {T.}~\bibnamefont {Wolf}},\ }\bibfield  {title} {\enquote {\bibinfo {title} {Absence of a second antiferromagnetic transition in pure {$\mathrm{Y}{\mathrm{Ba}}_{2}{\mathrm{Cu}}_{3}{\mathrm{O}}_{6+x}$}},}\ }\href {\doibase 10.1103/PhysRevB.50.9688} {\bibfield  {journal} {\bibinfo  {journal} {Phys. Rev. B}\ }\textbf
  {\bibinfo {volume} {50}},\ \bibinfo {pages} {9688--9691} (\bibinfo {year} {1994})}\BibitemShut {NoStop}%
\bibitem [{\citenamefont {Chaplot}\ \emph {et~al.}(1995)\citenamefont {Chaplot}, \citenamefont {Reichardt}, \citenamefont {Pintschovius},\ and\ \citenamefont {Pyka}}]{YBCO6phonon_expt}%
  \BibitemOpen
  \bibfield  {author} {\bibinfo {author} {\bibfnamefont {S.}~\bibnamefont {Chaplot}}, \bibinfo {author} {\bibfnamefont {W.}~\bibnamefont {Reichardt}}, \bibinfo {author} {\bibfnamefont {L.}~\bibnamefont {Pintschovius}}, \ and\ \bibinfo {author} {\bibfnamefont {N.}~\bibnamefont {Pyka}},\ }\bibfield  {title} {\enquote {\bibinfo {title} {Common interatomic potential model for the lattice dynamics of several cuprates},}\ }\href@noop {} {\bibfield  {journal} {\bibinfo  {journal} {Physical Review B}\ }\textbf {\bibinfo {volume} {52}},\ \bibinfo {pages} {7230} (\bibinfo {year} {1995})}\BibitemShut {NoStop}%
\bibitem [{\citenamefont {Wexler}, \citenamefont {Gautam},\ and\ \citenamefont {Carter}(2020)}]{PBESCANU}%
  \BibitemOpen
  \bibfield  {author} {\bibinfo {author} {\bibfnamefont {R.~B.}\ \bibnamefont {Wexler}}, \bibinfo {author} {\bibfnamefont {G.~S.}\ \bibnamefont {Gautam}}, \ and\ \bibinfo {author} {\bibfnamefont {E.~A.}\ \bibnamefont {Carter}},\ }\bibfield  {title} {\enquote {\bibinfo {title} {Exchange-correlation functional challenges in modeling quaternary chalcogenides},}\ }\href {\doibase 10.1103/PhysRevB.102.054101} {\bibfield  {journal} {\bibinfo  {journal} {Phys. Rev. B}\ }\textbf {\bibinfo {volume} {102}},\ \bibinfo {pages} {054101} (\bibinfo {year} {2020})}\BibitemShut {NoStop}%
\bibitem [{\citenamefont {Kothakonda}\ \emph {et~al.}(2022)\citenamefont {Kothakonda}, \citenamefont {Kaplan}, \citenamefont {Isaacs}, \citenamefont {Bartel}, \citenamefont {Furness}, \citenamefont {Ning}, \citenamefont {Wolverton}, \citenamefont {Perdew},\ and\ \citenamefont {Sun}}]{Manish_ACS}%
  \BibitemOpen
  \bibfield  {author} {\bibinfo {author} {\bibfnamefont {M.}~\bibnamefont {Kothakonda}}, \bibinfo {author} {\bibfnamefont {A.~D.}\ \bibnamefont {Kaplan}}, \bibinfo {author} {\bibfnamefont {E.~B.}\ \bibnamefont {Isaacs}}, \bibinfo {author} {\bibfnamefont {C.~J.}\ \bibnamefont {Bartel}}, \bibinfo {author} {\bibfnamefont {J.~W.}\ \bibnamefont {Furness}}, \bibinfo {author} {\bibfnamefont {J.}~\bibnamefont {Ning}}, \bibinfo {author} {\bibfnamefont {C.}~\bibnamefont {Wolverton}}, \bibinfo {author} {\bibfnamefont {J.~P.}\ \bibnamefont {Perdew}}, \ and\ \bibinfo {author} {\bibfnamefont {J.}~\bibnamefont {Sun}},\ }\bibfield  {title} {\enquote {\bibinfo {title} {Testing the r2scan density functional for the thermodynamic stability of solids with and without a van der waals correction},}\ }\href@noop {} {\bibfield  {journal} {\bibinfo  {journal} {ACS Materials Au}\ }\textbf {\bibinfo {volume} {3}},\ \bibinfo {pages} {102--111} (\bibinfo {year} {2022})}\BibitemShut {NoStop}%
\bibitem [{\citenamefont {Peng}\ and\ \citenamefont {Perdew}(2017{\natexlab{b}})}]{PBE_rVV10}%
  \BibitemOpen
  \bibfield  {author} {\bibinfo {author} {\bibfnamefont {H.}~\bibnamefont {Peng}}\ and\ \bibinfo {author} {\bibfnamefont {J.~P.}\ \bibnamefont {Perdew}},\ }\bibfield  {title} {\enquote {\bibinfo {title} {Rehabilitation of the perdew-burke-ernzerhof generalized gradient approximation for layered materials},}\ }\href {\doibase 10.1103/PhysRevB.95.081105} {\bibfield  {journal} {\bibinfo  {journal} {Phys. Rev. B}\ }\textbf {\bibinfo {volume} {95}},\ \bibinfo {pages} {081105} (\bibinfo {year} {2017}{\natexlab{b}})}\BibitemShut {NoStop}%
\bibitem [{\citenamefont {Falter}\ \emph {et~al.}(1999)\citenamefont {Falter}, \citenamefont {Klenner}, \citenamefont {Hoffmann},\ and\ \citenamefont {Schnetg\"oke}}]{model_Falter2}%
  \BibitemOpen
  \bibfield  {author} {\bibinfo {author} {\bibfnamefont {C.}~\bibnamefont {Falter}}, \bibinfo {author} {\bibfnamefont {M.}~\bibnamefont {Klenner}}, \bibinfo {author} {\bibfnamefont {G.~A.}\ \bibnamefont {Hoffmann}}, \ and\ \bibinfo {author} {\bibfnamefont {F.}~\bibnamefont {Schnetg\"oke}},\ }\bibfield  {title} {\enquote {\bibinfo {title} {Dipole polarization and charge transfer effects in the lattice dynamics and dielectric properties of ionic crystals},}\ }\href {\doibase 10.1103/PhysRevB.60.12051} {\bibfield  {journal} {\bibinfo  {journal} {Phys. Rev. B}\ }\textbf {\bibinfo {volume} {60}},\ \bibinfo {pages} {12051--12060} (\bibinfo {year} {1999})}\BibitemShut {NoStop}%
\bibitem [{\citenamefont {Falter}(2005)}]{model_Falter3}%
  \BibitemOpen
  \bibfield  {author} {\bibinfo {author} {\bibfnamefont {C.}~\bibnamefont {Falter}},\ }\bibfield  {title} {\enquote {\bibinfo {title} {Phonons, electronic charge response and electron--phonon interaction in the high-temperature superconductors},}\ }\href@noop {} {\bibfield  {journal} {\bibinfo  {journal} {physica status solidi (b)}\ }\textbf {\bibinfo {volume} {242}},\ \bibinfo {pages} {78--117} (\bibinfo {year} {2005})}\BibitemShut {NoStop}%
\bibitem [{\citenamefont {Kaplan}\ and\ \citenamefont {Perdew}(2022)}]{Aaron_OFR2}%
  \BibitemOpen
  \bibfield  {author} {\bibinfo {author} {\bibfnamefont {A.~D.}\ \bibnamefont {Kaplan}}\ and\ \bibinfo {author} {\bibfnamefont {J.~P.}\ \bibnamefont {Perdew}},\ }\bibfield  {title} {\enquote {\bibinfo {title} {Laplacian-level meta-generalized gradient approximation for solid and liquid metals},}\ }\href {\doibase 10.1103/PhysRevMaterials.6.083803} {\bibfield  {journal} {\bibinfo  {journal} {Phys. Rev. Mater.}\ }\textbf {\bibinfo {volume} {6}},\ \bibinfo {pages} {083803} (\bibinfo {year} {2022})}\BibitemShut {NoStop}%
\bibitem [{\citenamefont {Pintschovius}\ \emph {et~al.}(2004)\citenamefont {Pintschovius}, \citenamefont {Reznik}, \citenamefont {Reichardt}, \citenamefont {Endoh}, \citenamefont {Hiraka}, \citenamefont {Tranquada}, \citenamefont {Uchiyama}, \citenamefont {Masui},\ and\ \citenamefont {Tajima}}]{YBCO7_anomaly}%
  \BibitemOpen
  \bibfield  {author} {\bibinfo {author} {\bibfnamefont {L.}~\bibnamefont {Pintschovius}}, \bibinfo {author} {\bibfnamefont {D.}~\bibnamefont {Reznik}}, \bibinfo {author} {\bibfnamefont {W.}~\bibnamefont {Reichardt}}, \bibinfo {author} {\bibfnamefont {Y.}~\bibnamefont {Endoh}}, \bibinfo {author} {\bibfnamefont {H.}~\bibnamefont {Hiraka}}, \bibinfo {author} {\bibfnamefont {J.~M.}\ \bibnamefont {Tranquada}}, \bibinfo {author} {\bibfnamefont {H.}~\bibnamefont {Uchiyama}}, \bibinfo {author} {\bibfnamefont {T.}~\bibnamefont {Masui}}, \ and\ \bibinfo {author} {\bibfnamefont {S.}~\bibnamefont {Tajima}},\ }\bibfield  {title} {\enquote {\bibinfo {title} {Oxygen phonon branches in {${\mathrm{YBa}}_{2}{\mathrm{Cu}}_{3}{\mathrm{O}}_{7}$}},}\ }\href {\doibase 10.1103/PhysRevB.69.214506} {\bibfield  {journal} {\bibinfo  {journal} {Phys. Rev. B}\ }\textbf {\bibinfo {volume} {69}},\ \bibinfo {pages} {214506} (\bibinfo {year} {2004})}\BibitemShut {NoStop}%
\bibitem [{\citenamefont {Jana}, \citenamefont {Patra},\ and\ \citenamefont {Samal}(2018)}]{TM_metals}%
  \BibitemOpen
  \bibfield  {author} {\bibinfo {author} {\bibfnamefont {S.}~\bibnamefont {Jana}}, \bibinfo {author} {\bibfnamefont {A.}~\bibnamefont {Patra}}, \ and\ \bibinfo {author} {\bibfnamefont {P.}~\bibnamefont {Samal}},\ }\bibfield  {title} {\enquote {\bibinfo {title} {{Assessing the performance of the Tao-Mo semilocal density functional in the projector-augmented-wave method}},}\ }\href {\doibase 10.1063/1.5040786} {\bibfield  {journal} {\bibinfo  {journal} {The Journal of Chemical Physics}\ }\textbf {\bibinfo {volume} {149}},\ \bibinfo {pages} {044120} (\bibinfo {year} {2018})}\BibitemShut {NoStop}%
\bibitem [{\citenamefont {Juraschek}\ \emph {et~al.}(2017)\citenamefont {Juraschek}, \citenamefont {Fechner}, \citenamefont {Balatsky},\ and\ \citenamefont {Spaldin}}]{dyn_multiferro}%
  \BibitemOpen
  \bibfield  {author} {\bibinfo {author} {\bibfnamefont {D.~M.}\ \bibnamefont {Juraschek}}, \bibinfo {author} {\bibfnamefont {M.}~\bibnamefont {Fechner}}, \bibinfo {author} {\bibfnamefont {A.~V.}\ \bibnamefont {Balatsky}}, \ and\ \bibinfo {author} {\bibfnamefont {N.~A.}\ \bibnamefont {Spaldin}},\ }\bibfield  {title} {\enquote {\bibinfo {title} {Dynamical multiferroicity},}\ }\href {\doibase 10.1103/PhysRevMaterials.1.014401} {\bibfield  {journal} {\bibinfo  {journal} {Phys. Rev. Mater.}\ }\textbf {\bibinfo {volume} {1}},\ \bibinfo {pages} {014401} (\bibinfo {year} {2017})}\BibitemShut {NoStop}%
\bibitem [{\citenamefont {de~la Torre}\ \emph {et~al.}(2021{\natexlab{a}})\citenamefont {de~la Torre}, \citenamefont {Seyler}, \citenamefont {Zhao}, \citenamefont {Di~Matteo}, \citenamefont {Scheurer}, \citenamefont {Li}, \citenamefont {Yu}, \citenamefont {Greven}, \citenamefont {Sachdev}, \citenamefont {Norman} \emph {et~al.}}]{de2021}%
  \BibitemOpen
  \bibfield  {author} {\bibinfo {author} {\bibfnamefont {A.}~\bibnamefont {de~la Torre}}, \bibinfo {author} {\bibfnamefont {K.}~\bibnamefont {Seyler}}, \bibinfo {author} {\bibfnamefont {L.}~\bibnamefont {Zhao}}, \bibinfo {author} {\bibfnamefont {S.}~\bibnamefont {Di~Matteo}}, \bibinfo {author} {\bibfnamefont {M.}~\bibnamefont {Scheurer}}, \bibinfo {author} {\bibfnamefont {Y.}~\bibnamefont {Li}}, \bibinfo {author} {\bibfnamefont {B.}~\bibnamefont {Yu}}, \bibinfo {author} {\bibfnamefont {M.}~\bibnamefont {Greven}}, \bibinfo {author} {\bibfnamefont {S.}~\bibnamefont {Sachdev}}, \bibinfo {author} {\bibfnamefont {M.}~\bibnamefont {Norman}},  \emph {et~al.},\ }\bibfield  {title} {\enquote {\bibinfo {title} {Mirror symmetry breaking in a model insulating cuprate},}\ }\href@noop {} {\bibfield  {journal} {\bibinfo  {journal} {Nature Physics}\ }\textbf {\bibinfo {volume} {17}},\ \bibinfo {pages} {777--781} (\bibinfo {year} {2021}{\natexlab{a}})}\BibitemShut {NoStop}%
\bibitem [{\citenamefont {de~la Torre}\ \emph {et~al.}(2021{\natexlab{b}})\citenamefont {de~la Torre}, \citenamefont {Di~Matteo}, \citenamefont {Hsieh},\ and\ \citenamefont {Norman}}]{de2021b}%
  \BibitemOpen
  \bibfield  {author} {\bibinfo {author} {\bibfnamefont {A.}~\bibnamefont {de~la Torre}}, \bibinfo {author} {\bibfnamefont {S.}~\bibnamefont {Di~Matteo}}, \bibinfo {author} {\bibfnamefont {D.}~\bibnamefont {Hsieh}}, \ and\ \bibinfo {author} {\bibfnamefont {M.}~\bibnamefont {Norman}},\ }\bibfield  {title} {\enquote {\bibinfo {title} {Implications of second harmonic generation for hidden order in sr 2 cuo 2 cl 2},}\ }\href@noop {} {\bibfield  {journal} {\bibinfo  {journal} {Physical Review B}\ }\textbf {\bibinfo {volume} {104}},\ \bibinfo {pages} {035138} (\bibinfo {year} {2021}{\natexlab{b}})}\BibitemShut {NoStop}%
\bibitem [{\citenamefont {Ahmadova}\ \emph {et~al.}(2020)\citenamefont {Ahmadova}, \citenamefont {Sterling}, \citenamefont {Sokolik}, \citenamefont {Abernathy}, \citenamefont {Greven},\ and\ \citenamefont {Reznik}}]{Ahmadova2020}%
  \BibitemOpen
  \bibfield  {author} {\bibinfo {author} {\bibfnamefont {I.}~\bibnamefont {Ahmadova}}, \bibinfo {author} {\bibfnamefont {T.~C.}\ \bibnamefont {Sterling}}, \bibinfo {author} {\bibfnamefont {A.~C.}\ \bibnamefont {Sokolik}}, \bibinfo {author} {\bibfnamefont {D.~L.}\ \bibnamefont {Abernathy}}, \bibinfo {author} {\bibfnamefont {M.}~\bibnamefont {Greven}}, \ and\ \bibinfo {author} {\bibfnamefont {D.}~\bibnamefont {Reznik}},\ }\bibfield  {title} {\enquote {\bibinfo {title} {Phonon spectrum of underdoped {${\mathrm{HgBa}}_{2}{\mathrm{CuO}}_{4+\ensuremath{\delta}}$} investigated by neutron scattering},}\ }\href {\doibase 10.1103/PhysRevB.101.184508} {\bibfield  {journal} {\bibinfo  {journal} {Phys. Rev. B}\ }\textbf {\bibinfo {volume} {101}},\ \bibinfo {pages} {184508} (\bibinfo {year} {2020})}\BibitemShut {NoStop}%
\bibitem [{\citenamefont {Curtis}\ \emph {et~al.}(2022)\citenamefont {Curtis}, \citenamefont {Grankin}, \citenamefont {Poniatowski}, \citenamefont {Galitski}, \citenamefont {Narang},\ and\ \citenamefont {Demler}}]{cavity-phonon-magnon}%
  \BibitemOpen
  \bibfield  {author} {\bibinfo {author} {\bibfnamefont {J.~B.}\ \bibnamefont {Curtis}}, \bibinfo {author} {\bibfnamefont {A.}~\bibnamefont {Grankin}}, \bibinfo {author} {\bibfnamefont {N.~R.}\ \bibnamefont {Poniatowski}}, \bibinfo {author} {\bibfnamefont {V.~M.}\ \bibnamefont {Galitski}}, \bibinfo {author} {\bibfnamefont {P.}~\bibnamefont {Narang}}, \ and\ \bibinfo {author} {\bibfnamefont {E.}~\bibnamefont {Demler}},\ }\bibfield  {title} {\enquote {\bibinfo {title} {Cavity magnon-polaritons in cuprate parent compounds},}\ }\href {\doibase 10.1103/PhysRevResearch.4.013101} {\bibfield  {journal} {\bibinfo  {journal} {Phys. Rev. Research}\ }\textbf {\bibinfo {volume} {4}},\ \bibinfo {pages} {013101} (\bibinfo {year} {2022})}\BibitemShut {NoStop}%
\end{thebibliography}%

\end{document}